\documentclass[12pt,french,epsf,psfig]{elsart}
\usepackage[latin1]{inputenc}
\usepackage[T1]{fontenc}
\usepackage{fancyhdr}
\usepackage{amsmath}
\usepackage{graphics}
\usepackage{subfigure}
\usepackage[final]{epsfig}
\usepackage{mathrsfs}
\usepackage{amssymb}
\usepackage{multirow}
\usepackage{rotating}
\makeatletter

\newcommand{\LyX}{L\kern-.1667em\lower.25em\hbox{Y}\kern-.125emX\spacefactor1000}
\newcommand{\eq}[1]{Eq.~\ref{eq:#1}}
\newcommand{\fig}[1]{Fig.~\ref{fig:#1}}

\newcommand{\para}[1]{Sec.~\ref{para:#1}}
\newcommand{\ap}[1]{Appendix~\ref{ap:#1}}
\makeatother

\begin{document}
\begin{frontmatter}
 
\title{A Boundary-Based Net Exchange Monte-Carlo Method for absorbing and scattering thick media}

\maketitle
\begin{center}
\author[Toulouse]{V. Eymet\corauthref{corres}}
\author[Toulouse]{R. Fournier}
\author[Toulouse]{S. Blanco}
\author[Paris]{J.L. Dufresne}
\corauth[corres]{Corresponding author. Present address : Laboratoire d'Energétique, UFR-PCA, Université Paul Sabatier, 118 route de Narbonne, 31062 Toulouse Cedex, FRANCE \newline {\it E-mail address:} eymet@sphinx.ups-tlse.fr  (V. Eymet)}
\address[Toulouse]{Laboratoire d'Energétique, Université Paul Sabatier, 31062 Toulouse, France}
\address[Paris]{LMD/CNRS, Université de Paris VI, 75252 Paris, France}
\end{center}

\begin{abstract}
A boundary-based net-exchange Monte Carlo method was introduced in \cite{amaury02} that allows to bypass the difficulties encountered by standard Monte Carlo algorithms in the limit of optically thick absorption (and/or for quasi-isothermal configurations). With the present paper, this method is extended to scattering media. Developments are fully 3D, but illustrations are presented for plane parallel configuration. Compared to standard Monte Carlo algorithms, convergence qualities have been improved over a wide range of absorption and scattering optical thicknesses. The proposed algorithm still encounters a convergence difficulty in the case of optically thick, highly scattering media.
\end{abstract}
\begin{keyword}
Monte Carlo, net-exchange formulation, scattering, numerical optimization, convergence
\end{keyword}
\end{frontmatter}

\section{Introduction}

The Monte-Carlo Method (MCM) has been widely used in the field of transport phenomena simulation, and more specifically in the field of radiative transfer computing \cite{Hammersley,Howell02,Howell01}. In this particular case, the method mainly consists in simulating numerically the physical statistical model of photons transport, from their emission to their absorption in a potentially scattering medium. A well known advantage of this method is that the corresponding computing code is easy to set up and to modify. Another main advantage is that it is a reference method~: as the MCM is a statistical method, a standard deviation may be computed in addition to each result, that may be interpreted as a numerical uncertainty. Also, it has recently been shown that the MCM allows the computation of parametric sensitivities with no extra significant computing \cite{amaury01}. This can be helpful for design needs, or when radiative transfer is coupled with other physical processes. Finally, the MCM is known to be well adapted to the treatment of configurations with a high level of complexity (complex geometries, complex spectral properties, ...). However, in spite of these advantages over other methods and in spite of the regular increase of available computational powers, the computational effort requirement of MCM often remains a significant drawback.

Different works in the last fifteen years tried to preserve the main advantages of the method, in particular its strict analogy with physical processes, and the ability to solve complex problems, while trying to improve convergence qualities. There are mainly two ways MCM convergence can be enhanced~: formulation changes and adaptation of sampling laws \cite{Hammersley}. As far as formulation is concerned, most attention has been devoted to reverse Monte Carlo algorithms \cite{Walters}, that make use of reciprocal transport formulations (application of the reciprocity principle to the integral form of the radiative transfer equation), and to net-exchange Monte Carlo algorithms \cite{Fournier04,Fournier03,Dufresne00,Tesse01}, that make use of net exchange transport formulations (combination of forward and reciprocal formulations, photons being followed both ways along each optical path). Net-exchange Monte Carlo algorithms allowed in particular to bypass the problem of standard Monte Carlo algorithms for quasi-isothermal configurations. As far as sampling laws optimization is concerned, numerous works have successfully used the biasing of sampled directions toward the parts of the system that most contribute to the addressed radiative quantity \cite{Martin}, or the biasing of sampled frequencies as function of temperature field and spectral properties \cite{Fournier04,Dufresne01}.

Recently, the combination of formulation efforts and sampling laws adaptations permitted to solve the well known convergence problem of traditional Monte-Carlo algorithms in the case of strong optical thickness configurations \cite{amaury02}. If a gas volume is optically very thick, most emitted photons are absorbed very close to their emission position, and thus do not take part to the exchange of the gas volume with the rest of the system. Consequently, very large numbers of statistical realizations are required to reach satisfactory convergences. This problem could be solved in the case of purely absorbing systems thanks to a net-exchange formulation in which emission positions are sampled, starting form the volume boundary, along an inward oriented sampled direction (a formulation that will be named here a "boundary-based net-exchange formulation"). All sampling laws (frequency, boundary position, direction and emission position) where also finely optimized in order to insure that the algorithm automatically adapts to system optical thickness in the whole range from the optically thin to the optically thick limits.

The present paper is one of a series that seek to improve the MCM through such methodological developments. It proposes techniques to take into account scattering in the above mentioned boundary-based net-exchange algorithm. The formulation used in \cite{amaury02} has been generalized and clarified in order to take into account the scattering phenomena. Developments are fully 3D, but convergence illustrations are presented for plane parallel configurations that are specifically meaningful in the atmospheric science community.

\para{part1} of this article puts the emphasis on the multiple integral theoretical developments on which our Monte-Carlo algorithm is based. \para{part2} presents gas volume emission results in a simple test case, thus revealing the algorithm convergence qualities together with its limits of applicability. Finally, \para{part3} completes this convergence illustration in terms of radiative flux divergence profiles.

\section{Theoretical developments}
\label{para:part1}

The three next paragraphs deal with improvements that were brought to the standard bundle transport MCM during the last few years, through a number of different methodological developments.

\subsection{Exchange Formulation}

Let us consider that, for the purpose of a 3D radiative transfer computation,
the considered system is divided into volume and surface elements. Until
further mention, this geometric division is only motivated by the required
level of analysis and it therefore implies no physical assumption~: the
volume and surface elements have any geometrical shapes and are inhomogeneous.

\begin{figure}[h!t]
  \begin{center}
        \epsfig{figure=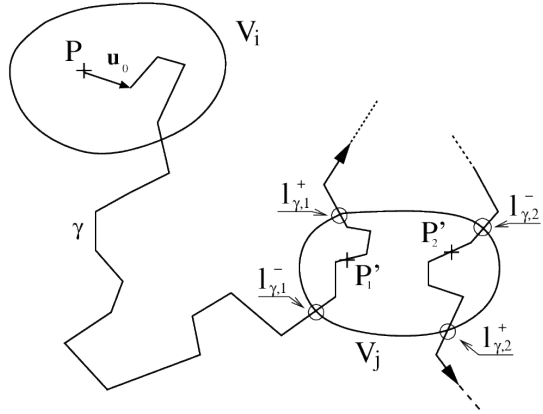 ,width=0.65\textwidth}
    \caption{Discretization of an absorbing and scattering semi-transparent medium into volume elements}
    \label{fig:raytracing}
  \end{center}
\end{figure}

The energy rate $E_{i,j}$ emitted by an arbitrary gas volume $i$ and absorbed by an arbitrary gas volume $j$ may be expressed as~:
\begin{equation}
\begin{split}
E_{ij}= & \int_{V_{i}}dV_{i}(P)\int_{4\pi}d\omega({\bf u_{0}})\int_{\Gamma_{(P,{\bf u_{0}})}} p(\gamma;P,{\bf u_{0}}) d\gamma \ k_{a}(P)\sum_{n=1}^{\infty} T_{\gamma,n}\int_{l_{\gamma,n}^{-}}^{l_{\gamma,n}^{+}} d\sigma_{n}(P_n^{\prime}) \\
& k_{a}(P_n^{\prime}) B(P) exp\Bigl(-\int_{l_{\gamma,n}^{-}}^{\sigma_{n}}d\sigma^{\prime} k_{a}(\sigma^{\prime})\Bigr)
\end{split}
\label{eq:Eij}
\end{equation}
where $B(P)$ is the monochromatic blackbody intensity at point $P$. $\Gamma_{(P,{\bf u_{0}})}$ represents the space of (infinite) optical paths $\gamma$ originated from point $P$, in the direction ${\bf u_{0}}$, distributed according to $p(\gamma;P,{\bf u_{0}})$. Every such path will finally reach volume $V_{j}$ and will even cross it an infinite number of times. $l_{\gamma}$ is the curvilinear coordinate along the optical path $\gamma$ and $l_{\gamma,n}$ are the values of this curvilinear coordinate at the positions of the $n^{th}$ intersection between optical path $\gamma$ and gas volume $V_{j}$ : $l_{\gamma,n}^{-}$ stands for the $n^{th}$ entry point coordinate, and $l_{\gamma,n}^{+}$ stands for the $n^{th}$ exit point coordinate. $\sigma_{n}$ is the curvilinear abscissa of point $P_n^{\prime}$ in the $n^{th}$ intersection interval between $\gamma$ and $V_{j}$. $T_{\gamma,n}$ is the transmitivity between point $P$ and the position $l_{\gamma,n}^{-}$ : it is a product of exponential attenuations and of surface reflectivities for surface reflexions.

The integral over $\Gamma_{(P,{\bf u_{0}})}$, according to the distribution $p(\gamma;P,{\bf u_{0}})$, will not be detailed in this paper, because the purpose of this work is to put the emphasis on the Net Exchange Formulation itself, and not to deal with the physical model used for optical paths representation. These paths are purely random walk optical paths, and all the complexities associated with the formulation of scattering angles and free path length are deported into the expression of $p(\gamma;P,{\bf u_{0}})$, that represents formally the existence probability density of a given optical path $\gamma$.

This formulation may be used to derive a standard path integrated Monte-Carlo algorithm, that may be described as follows :
\begin{itemize}
\item First, the emission point $P$ is randomly chosen in the gas volume $V_{i}$, and the emission direction ${\bf u_{0}}$ is randomly chosen in the unit sphere
($4\pi$ st).
\item The optical path $\gamma$ is generated with a standard random walk technique.
\item Each time this optical path reaches the gas volume $V_{j}$, a point $P^{\prime}$ is randomly chosen along the part of $\gamma$ that intersects $V_{j}$.
\item Finally, the optical path $\gamma$ ends when it is long enough for the energy bundle to be considered as totally attenuated (as function of the required level of accuracy).
\end{itemize}

Using a Monte-Carlo algorithm based on a traditional formulation, the Net Radiative Budget is expressed as the difference between approximate emitted and absorbed energy rates that are computed separately, and that can be close the one to the other for nearly isothermal configurations, inducing numerical convergence difficulties.

\subsection{Net Exchange Formulation}

The Net Exchange Formulation is based on the Net Exchange Rates $\Psi_{ij}$ between all pairs of elements (either surface or volume elements) $i$ and $j$, which is defined as the difference between the energy rate emitted from element $i$ and absorbed by element $j$, and the energy emitted from element $j$ and absorbed by element $i$.

The advantages of formulating radiative transfers in terms of Net Exchange Rates have been shown by Green \cite{Green}. The Net Exchange Formulation has been introduced in the MCM by \cite{Fournier04,Fournier03}. This radiative transfer formulation solved the convergence problem encountered by the MCM in nearly isothermal configurations. A general formulation of a Net Exchange Rate $\Psi_{ij}$ between two gas volumes $i$ and $j$ may be directly deduced from the energy rate equation \eq{Eij}, replacing $B(P)$ by $\Bigl[B(P)-B(P^{\prime})\Bigr]$~\cite{Dufresne00}~:
\begin{equation}
\begin{split}
\Psi_{ij}= & \int_{V_{i}}dV_{i}(P)\int_{4\pi}d\omega({\bf u_{0}})\int_{\Gamma_{(P,{\bf u_{0}})}} p(\gamma;P,{\bf u_{0}}) d\gamma \ k_{a}(P)\sum_{n=1}^{\infty} T_{\gamma,n}\int_{l_{\gamma,n}^{-}}^{_{\gamma,n}^{+}} d\sigma_{n}(P_n^{\prime}) \\
& k_{a}(P_n^{\prime})\Bigl[B(P)-B(P_n^{\prime})\Bigr]exp\Bigl(-\int_{l_{\gamma,n}^{-}}^{\sigma_{n}}d\sigma^{\prime} k_{a}(\sigma^{\prime})\Bigr)
\end{split}
\label{eq:pne1}
\end{equation}
where $B(P_n^{\prime})$ is the monochromatic blackbody intensity at point $P_n^{\prime}$ in volume $V_{j}$ (see \fig{raytracing}). Similarity of equations \eq{Eij} and \eq{pne1} makes fairly easy the implementation of a Net Exchange Formulation in a standard Monte Carlo algorithm~: the monochromatic blackbody intensity $B(P)$ has to be replaced by $\Bigl[B(P)-B(P^{\prime})\Bigr]$.

Using a Net Exchange Formulation, the Net Radiative Budget of a gas volume $i$ may be expressed as a sum of Net Exchange Rates~:
\begin{equation}
\Psi_{i}=\sum_{j} \Psi_{ij}
\label{eq:bilan}
\end{equation}

Here, using a Monte-Carlo algorithm based on a Net Exchange Formulation means that all Net Exchange Rates are computed separately (as pondered sums of blackbody intensity differences $\Bigl[B(P)-B(P^{\prime})\Bigr]$, which induces no numerical difficulty) and are then added to produce the Net Radiative Budget. In the limit case of nearly isothermal configurations, no convergence difficulty will be encountered~: as it does no longer compute the difference between two very close approximate values, a Monte-Carlo algorithm based on a Net Exchange Formulation will lead to much better accuracies than traditional Monte-Carlo algorithms.\cite{Fournier04,Fournier03}.

\subsection{Boundary-Based Net Exchange Formulation}

\begin{figure}[h!t]
  \begin{center}
        \epsfig{figure=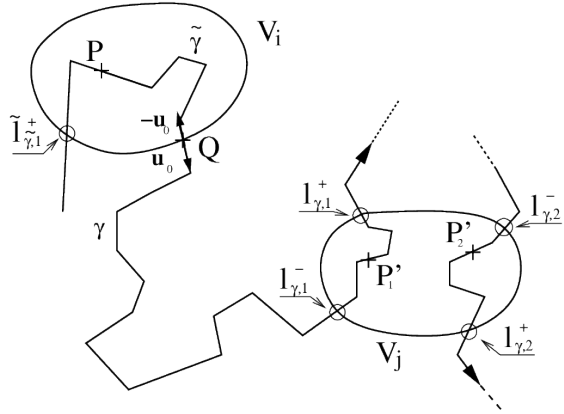 ,width=0.65\textwidth}
    \caption{Boundary-based reformulation of net exchange rates}
    \label{fig:raytracing2}
  \end{center}
\end{figure}

A typical difficulty that is encountered by any standard MCM (both bundle transport and path integrated MC algorithms) \footnote{The term ``bundle transport algorithm'' is used for algorithms in which the photon bundle's energy is totally absorbed at a stochastically determined position. The term ``path integrated algorithm'' is used for algorithms in which the photon bundle's energy is exponentially attenuated along the photon bundle's optical path} is the problem of optically thick systems. Let us consider the computation of the emission $E$ from a given gas volume, using a standard path integrated Monte-Carlo algorithm. In the case of an optically thick gas volume, the computation of $E$ will suffer from a convergence problem : most bundles emitted into the gas volume will be totally attenuated when they cross the volume boundary. Only those emitted very close to the boundary will have a chance to leave the gas volume with a significant computational weight. Thus, the computation of $E$ will require a great number of statistical realizations $N$ in order to get a good accuracy over $E$. This convergence difficulty is due to the fact that emission positions are chosen uniformly among the gas volume. A possible way to solve this problem would be to sample more often emission positions close to the volume boundary, so that most bundles would leave the gas volume with a significant energy, thus contributing more significantly to $E$. Modifying the way emission positions are sampled means to modify sampling laws used in the algorithm, without modifying the result of the multiple integral ; in order to do this, we choose to use a net exchange formulation different from the initial formulation presented in \eq{pne1}. This reformulation -~that brings forward the distance between emission point and first exit point~- is the purpose of the present subsection.

\eq{pne1} starts with an integration over all locations $P$ within volume $V_i$, then one integrates over all optical paths $\gamma$ starting at $P$ and $\gamma$ happens to cross the boundary of $V_i$ (here noted $S_i$) at a location $Q$ (see \fig{raytracing2})~: this boundary does not appear as an explicit integration domain. On the contrary, the following formulation (that will be referred as ``Boundary-Based Net Exchange Formulation'') starts with an integration over all exit locations $Q$ on $S_i$, then one integrates over the exit hemisphere at $Q$ and then over all the optical paths initiating within $V_i$ and crossing its boundary at the retained exit location and exit direction~: the boundary of $V_i$ appears as an explicit integration domain, but not the volume $V_i$ itself.
\begin{equation}
\begin{split}
\Psi_{ij}= & \int_{S_{i}}dS_{i}(Q)\int_{2\pi}d\omega({\bf u_{0}}) \ {\bf u_{0}}.{\bf n} \ L_{i}(Q,{\bf u_{0}})\int_{\Gamma_{(Q,{\bf u_{0}})}} p(\gamma;Q,{\bf u_{0}}) d\gamma \\
& \sum_{n=1}^{\infty} T_{\gamma,n}\int_{l_{\gamma,n}^{-}}^{l_{\gamma,n}^{+}} d\sigma_{n}(P_n^{\prime})k_{a}(P)k_{a}(P_n^{\prime})\Bigl[B(P)-B(P_n^{\prime})\Bigr]exp\Bigl(-\int_{l_{\gamma,n}^{-}}^{\sigma_{n}}d\sigma^{\prime} k_{a}(\sigma^{\prime})\Bigr)
\end{split}
\label{eq:pne2}
\end{equation}
where $L_{i}(Q,{\bf u_{0}})$ is the fraction of the intensity at $Q$ in direction ${\bf u_{0}}$ that corresponds to photons emitted within $V_{i}$ and crossing $S_i$ for the first time. Using the optical path reciprocity principle, it is possible to formulate $L_{i}(Q,{\bf u_{0}})$ as~:
\begin{equation}
L_{i}(Q,{\bf u_{0}})=\int_{\Gamma_{(Q,{\bf -u_{0}})}} p(\tilde{\gamma};Q,{\bf -u_{0}})  d\tilde{\gamma} \int_{0}^{\tilde{l}_{\tilde{\gamma},1}^{+}} d\tilde{\sigma}(P) \ exp\Bigl(-\int_{0}^{\tilde{\sigma}} d\tilde{\sigma^{\prime}} k_{a}(\tilde{\sigma^{\prime}})\Bigr)
\label{eq:Li1}
\end{equation}
where $\Gamma_{(Q,{\bf -u_{0}})}$ is the space of optical paths originated from point $Q$, in the direction ${\bf -u_{0}}$ and $\tilde{l}_{\tilde{\gamma},1}^{+}$ stands for the point at which $\tilde{\gamma}$ first exits $V_i$ (see \fig{raytracing2}).


The Monte-Carlo algorithm that was derived from this new formulation of Net Exchange Rates $\Psi_{ij}$ may be described as follows~:

\begin{itemize}
\item First, a point $Q$ is randomly chosen on the boundary $S_{i}$ surrounding gas volume $V_{i}$, and the initial direction ${\bf u_{0}}$ is randomly chosen in the exit unit hemisphere ($2\pi$ st).
\item Starting at $Q$ in the direction $-{\bf u_{0}}$, the optical path $\tilde{\gamma}$ is generated with a standard random walk technique until it first exits $V_{i}$ and $P$ is then randomly chosen within $V_{i}$ along this truncated path.
\item Starting at $Q$ in the direction ${\bf u_{0}}$, the optical path $\gamma$ is generated with a standard random walk technique.
\item Each time $\gamma$ reaches volume $V_{j}$, a point $P^{\prime}$ is randomly chosen along the part of $\gamma$ that intersects $V_{j}$.
\item Finally, the optical path $\gamma$ ends when it is long enough for the net-exchange bundle to be considered as totally attenuated (as function of the required level of accuracy).
\end{itemize}

At this point of the developments, only the boundary-based reformulation of net exchange rates has been achieved. In the next subsection, it will be shown how the sampling laws that arise from this formulation (Monte Carlo computation of the corresponding multiple integrals) may be optimized in order to solve convergence difficulties in the optically thick limits.

\subsection{Optimization of sampling laws}

The above described algorithm principle requires successive random generations\footnote{The random walk sampling laws corresponding to the generations of $\tilde{\gamma}$ and $\gamma$ are left apart in the present article because no optimization is proposed concerning this part of the algorithm. Such an optimization process is non trivial and none of the attempts made at date have enough generality to be implemented on a standard basis. Among the most successful attempts, a specific mention can be made to the work of Berger and al. reported in \cite{Hammersley} for simulation of optically thick radiation shields.} of an exit position $Q$, an exit direction ${\bf u_{0}}$, a first exchange position $P$ (via the curvilinear abscissa $\tilde{\sigma}$), and second exchange positions $P^{\prime}$ (via the curvilinear abscissa $\sigma$). It can be easily shown that any non zero probability density function may be used for each such sampling insuring the same integral solution at the limit of an infinite number of bundles. One way of illustrating this point is to rewrite Net Exchanges Rates $\Psi_{ij}$, starting from \eq{pne2} and transforming all successive integrals into statistical averages~:
\begin{equation}
\begin{split}
\Psi_{ij}= & <\sum_{n=1}^{\infty} I_n \ \frac{1}{\beta_n}> \\
= & \int_{S_{i}} pdf_{S}(Q) dS_{i}(Q)
\int_{2\pi} pdf_{\Omega}({\bf u_{0}}) d\omega({\bf u_{0}}) \\
& \int_{\Gamma_{(Q,{\bf -u_{0}})}} p(\tilde{\gamma};Q,{\bf -u_{0}}) d\tilde{\gamma}
\int_{0}^{\tilde{l}_{\tilde{\gamma},1}^{+}} pdf_{\tilde{\Sigma}}(\tilde{\sigma}) d\tilde{\sigma}(P)  \\
& \int_{\Gamma_{(Q,{\bf u_{0}})}} p(\gamma;Q,{\bf u_{0}}) d\gamma \ \prod_{n=1}^{\infty} 
\left(
\int_{l_{\gamma,n}^{-}}^{l_{\gamma,n}^{+}} pdf_{\Sigma_{n}}(\sigma_{n}) d\sigma_{n}(P^{\prime})
\right)
\left\{ \sum_{n=1}^{\infty} I_n \ \frac{1}{\beta_n} \right\}
\end{split}
\label{eq:pne3}
\end{equation}
where $I_n$ is the net-exchange density~:
\begin{equation}
I_n= ({\bf u_{0}}.{\bf n}) \  k_{a}(P) exp\Bigl(-\int_{0}^{\tilde{\sigma}}k_{a}(\tilde{\sigma^{\prime}})d\tilde{\sigma^{\prime}}\Bigr) \  T_{\gamma,n} \ k_{a}(P_n^{\prime})\Bigl[B(P)-B(P_n^{\prime})\Bigr]exp\Bigl(-\int_{l_{\gamma,n}^{-}}^{\sigma_{n}} d\sigma^{\prime} k_{a}(\sigma^{\prime}) \Bigr)
\label{eq:density}
\end{equation}
and $\beta$ the correction term :
\begin{equation}
\beta_n= pdf_{S}(Q) pdf_{\Omega}(\omega) pdf_{\tilde{\Sigma}}(\tilde{\sigma})  pdf_{\Sigma_{n}}(\sigma_{n})
\label{eq:alpha}
\end{equation}
\eq{pne3}, \eq{density} and \eq{alpha} insure a continuous link between the retained photon transport model (with a given formulation choice, here \eq{pne2}) and the Monte Carlo algorithm~: successive integral averages are translated into successive random sampling events and for each set of sampled variables the retained quantity is the sum of all $I_n  \frac{1}{\beta_n}$ (whose average value will be an approximation of $\Psi_{ij}$). Once the transport model and the integral formulation have been chosen the only remaining question is the choice of the sampling probability density functions~: this last choice does not modify the algorithmic structure, neither does it change the solution after convergence, but it strongly affects algorithmic convergence via the variance of $\sum_{n=1}^{\infty} I_n \ \frac{1}{\beta_n}$. The more physical knowledge is introduced in these probability density functions, the smaller the variance of $\sum_{n=1}^{\infty} I_n \ \frac{1}{\beta_n}$ and the faster the convergence\cite{Hammersley}. The probability density functions proposed hereafter are designed to insure satisfactory convergence speeds for a wide range of absorption and scattering optical thicknesses. The main objective was generality, hopping that such a set of probability density functions can serve as a start basis for more detailed adjustments when addressing specific configurations families.

\begin{itemize}

\item Sampling of exit points $Q$

The boundary sampling law $pdf_{S}(Q)$ has been chosen as uniform~: $pdf_{S}(Q)=1/S_i$. In the general case, having no information concerning the parts of $S_i$ through which $V_i$ exchanges most radiative energy with its environment, no better pdf adjustment could be proposed. Obviously, for specific configurations where $V_i$ exchanges radiation with hot spots at identified locations, this information can be directly used to modify $pdf_{S}(Q)$ so that the areas of stronger net-exchanges are more frequently sampled.

\item Sampling of exit directions ${\bf u_{0}}$

In the work of De Lataillade and al.(\cite{amaury02}), the angular sampling law $pdf_{\Omega}({\bf u_{0}})$ was optimized for the case of a purely absorbing medium. The lambertian distribution was used for strong optical thicknesses, whereas the isotropic distribution was used in case of optically thin gas volumes. The limit  between weak and strong optical thicknesses was set to $\tau_{a}=1$ where $\tau_{a}$ is the absorption optical thickness of the considered volume~:
\begin{equation}
\begin{split}
pdf_{\Omega}({\bf u_{0}})= & \frac{1}{2\pi} \ if \ \tau_{a}<1\\ 
pdf_{\Omega}({\bf u_{0}})= & \frac{{\bf u_{0}}.{\bf n}}{\pi} \ if \ \tau_{a} \ge 1
\end{split}
\label{eq:dir1}
\end{equation}
In the present work, the limit criteria is modified in order to account for the effect of scattering.
\begin{equation}
\begin{split}
pdf_{\Omega}({\bf u_{0}})= & \frac{1}{2\pi} \ if \ \tau_{eq}=\tau_{a}+(1-g)\tau_{s}<1\\ 
pdf_{\Omega}({\bf u_{0}})= & \frac{{\bf u_{0}}.{\bf n}}{\pi} \ if \ \tau_{eq}=\tau_{a}+(1-g)\tau_{s} \ge 1
\end{split}
\label{eq:dir2}
\end{equation}
where $\tau_{s}$ is the scattering optical thickness of the considered volume and $g$ is the phase function asymmetry parameter. In the case of a purely absorbing medium, $\tau_{eq}=\tau_{a}$ and we are back to the proposition of \cite{amaury02}~: when $\tau_{a}>1$, the absorption mean free path $\lambda_{a}=\frac{1}{k_a}$ is smaller than the system size which insures that the specific intensity of emitted photons (photons emitted within the gas volume that reach the boundary) is close to isotropy. Multiple scattering also induces an isotropic distribution of specific intensity at the volume boundary, but here the relevant scale is not the scattering mean free path $\lambda_{s}=\frac{1}{k_s}$ but the scattering {\it transport} mean free path $\frac{\lambda_{s}}{1-g}$ which accounts for the shape of the scattering phase function (forward scattering induces higher values of the transport mean free path)\cite{case}. When the medium both absorbs and scatters, the relevant scale is the total transport mean free path $\lambda_{eq}$ defined as $\frac{1}{\lambda_{eq}}=\frac{1}{\lambda_{a}}+\frac{1-g}{\lambda_{s}}$ which leads to the proposition of \eq{dir2}.

\item Sampling of first exchange position $P$

As in \cite{amaury02}, the first exchange position $P$ along $\tilde{\gamma}$ is sampled by use of a randomly generated abscissa $\tilde{\sigma}$ between $0$ and $\tilde{l}_{\tilde{\gamma},1}^{+}$ (see \fig{raytracing2}). The main interest of the proposed boundary based formulation is that the sampling law can be chosen as function of the absorption optical thickness in order to favor emission positions close to the boundary in the optically thick limit. This is done using an exponential probability density function for $\tilde{\gamma}$, which corresponds to an ideal adaptation for isothermal gas volumes~:
\begin{equation}
pdf_{\tilde{\Sigma}}(\tilde{\sigma})=
\frac{k_{a}exp(-k_{a}\tilde{\sigma})}{1-exp(-k_{a} \tilde{l}_{\tilde{\gamma},1}^{+})}
\label{eq:pdfP}
\end{equation}
Random generation of $\tilde{\gamma}$ is simply performed on the basis of a uniform random generation of $r$ in the unit interval according to~:
\begin{equation}
\tilde{\sigma}=-\frac{1}{k_{a}}ln\Biggl(1-r\Bigl(1-exp(-k_{a} \tilde{l}_{\tilde{\gamma},1}^{+})\Bigr)\Biggr)
\label{eq:length}
\end{equation}
For small values of absorption coefficient $k_{a}$ (optically thin limit), the above expression reduces to $\tilde{\sigma} \approx r \tilde{l}_{\tilde{\gamma},1}^{+}$, which is equivalent to choosing uniformly $\tilde{\sigma}$ within $[0,\tilde{l}_{\tilde{\gamma},1}^{+}]$. The physical significance of this, is that each point of $\tilde{l}_{\tilde{\gamma}}$ contributes the same way to the radiative transfer, because the energy emitted at each point is totally transmitted out of the gas volume.
On the contrary, for strong values of $k_{a}$ (optically thick limit), \eq{length} reduces to $\tilde{\sigma} \approx -\frac{1}{k_{a}}ln(1-r)$. $\tilde{l}_{\tilde{\gamma},1}^{+}$ is no longer taken into account and most exchange positions $P$ are sampled in the immediate vicinity of the boundary~: most statistical events have a significant contribution to the net-exchange and the statistical variance is reduced.

\item Sampling of second exchange positions $P^{\prime}_n$

Similarly, second exchange positions are generated along $\gamma$ by use of randomly generated abscissa $\sigma_n$ according to~:
\begin{equation}
pdf_{\Sigma_n}(\sigma_n)=
\frac{k_{a}exp(-k_{a}(\sigma_n - l_{\gamma,n}^{-}))}{1-exp(-k_{a} (l_{\gamma,n}^{+} - l_{\gamma,n}^{-}))}
\label{eq:pdfP2}
\end{equation}
\end{itemize}

Unlike in \cite{amaury02}, when the medium is both absorbing and scattering, the impact of these sampling laws on the behavior of the associated Monte-Carlo algorithm is configuration dependent~: sampling law adaptation is not satisfactory in the whole parameter range. The leading parameter is the single scattering albedo~: $\omega=\frac{k_{s}}{k_{a}+k_{s}}$

\begin{itemize}

\item For $\omega<<1$, scattering is negligible compared to absorption. In this case, the medium may be considered as purely absorbing, and it has been shown in \cite{amaury02} that the proposed sampling laws are suitable for such configurations. In particular, they solve the convergence difficulty encountered by Monte-Carlo algorithms in optically thick absorption configurations.

\item For usual values of $\omega$ ($\omega \in ]0,1[$ except for values very close to unity), scattering increases optical path lengths, and the use of the presented sampling laws results in a correct sampling of both exchange positions $P$ and $P^{\prime}$.

\item For $\omega \approx 1$, absorption is negligible compared to scattering. In this particular case, the proposed sampling laws fail to sample efficiently the optical path space. The difficulty may be described as follows : when scattering is the dominant process, the medium may be considered as optically thin on the point of view of absorption. In this case, all points into a given gas volume contribute equally to the exchange between this gas volume and the rest of the system. Even if the use of the proposed law for $pdf_{\tilde{\Sigma}}(\tilde{\sigma})$ will result in a uniform sampling of first exchange positions $P$  along all generated optical paths, most of these paths will be very short, because of the medium strong scattering properties (intense backscattering from point $Q$). First exchange positions $P$ will therefore be mainly sampled in the vicinity of the volume boundary which is not in accordance with the physics of radiative net-exchanges in little absorbing and highly scattering configurations. The proposed algorithm will therefore encounter convergence difficulties. We will see however that this difficulty is partly compensated by a reduction of the average number of scattering events to be numerically generated, the overall cost of the algorithm remaining satisfactory up to high albedo levels.

\end{itemize}

\section{Convergence illustration~: non-isothermal slab emission}
\label{para:part2}

As in \cite{amaury02}, the proposed algorithm is first tested using the academic problem of monochromatic slab emission. A single horizontal slab is considered, constituted of semi-transparent medium, with uniform absorbing and scattering optical properties, between two black boundaries at 0K. The slab physical thickness is $H$ and the z-axis is downward-positive. The temperature profile across the slab is such that the blackbody intensity at the considered frequency follows a linear profile $B(z)$ from $0$ at the top to $B_{0}$ at the bottom of the slab. The addressed quantity is the downward slab emission, which is also the net-exchange rate between the slab and the bottom boundary. 

\begin{figure}[htbp]
\centering
\mbox{
    \subfigure[$N$ for $\omega=0.01$]{\epsfig{figure=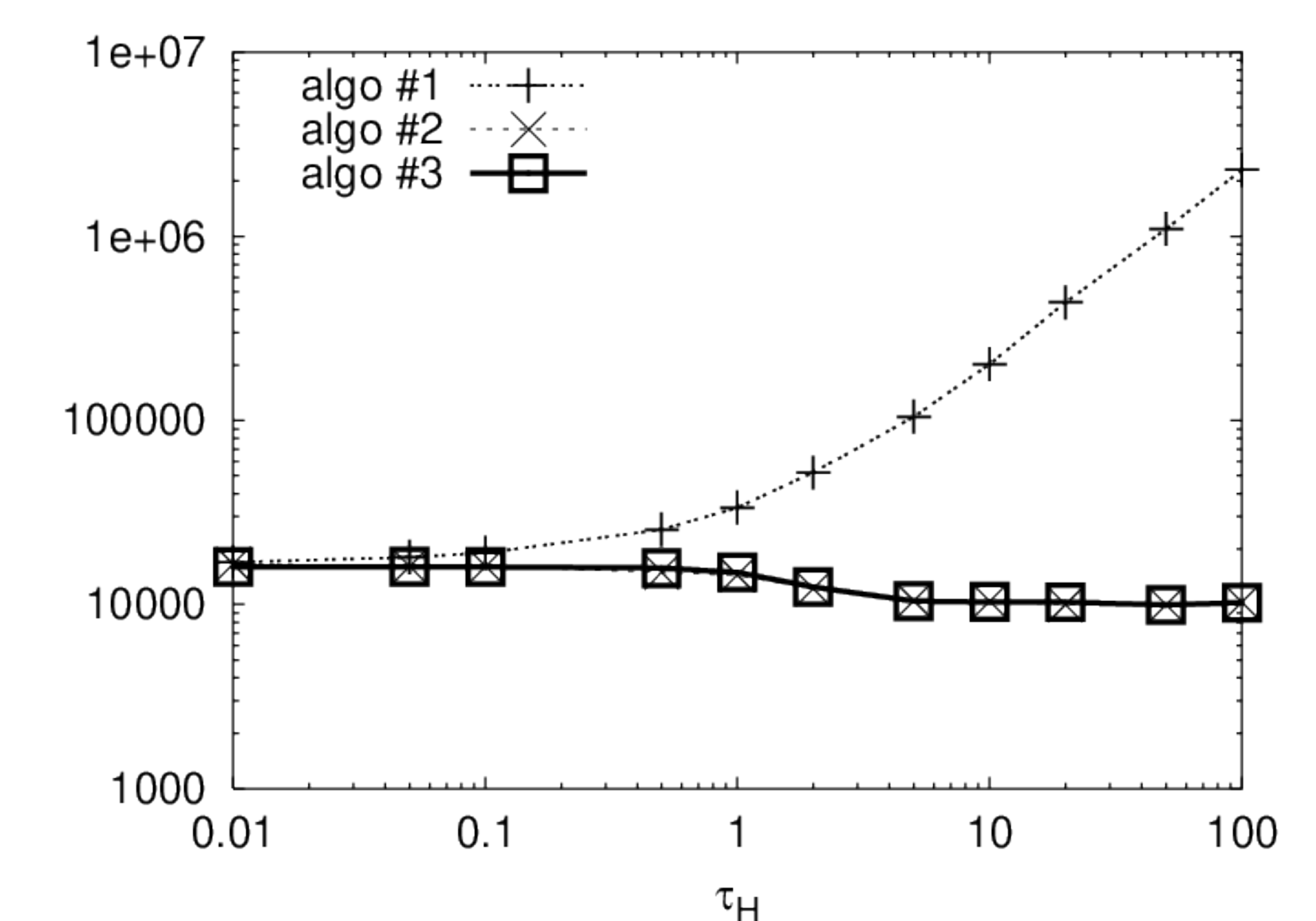,width=0.40\textwidth}}\quad
    \subfigure[$<N_{s}>$ for $\omega=0.01$]{\epsfig{figure=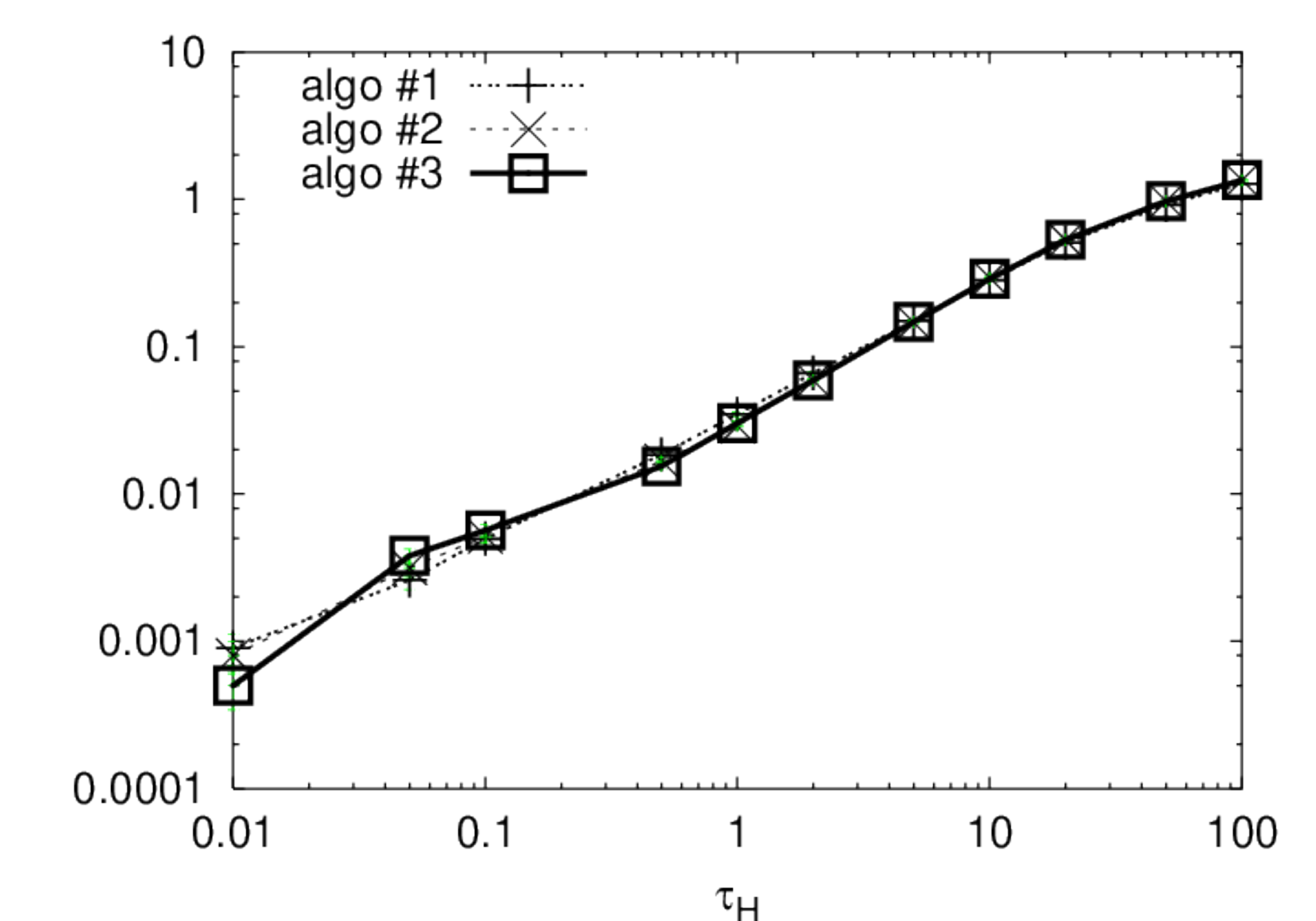,width=0.40\textwidth}}}
    \caption{(a)~: Number of statistical realizations $N$ required to compute slab emission with a relative standard deviation of $1$ percent as function of slab total optical thickness $\tau_{H}$. (b)~: Average number of scattering events $<N_{s}>$ as function of slab total optical thickness $\tau_{H}$. Calculations held with $\omega=0.01$. Presented results correspond to three different algorithms~: standard Monte Carlo algorithm (algo \#1), boundary-based net-exchange algorithm (algo \#2), boundary-based net-exchange algorithm without the optimization of angular sampling as function of scattering (algo \#3).}
\label{fig:conv1}
\end{figure}

\begin{figure}[htbp]
\centering
\mbox{
    \subfigure[$N$ for $\omega=0.50$]{\epsfig{figure=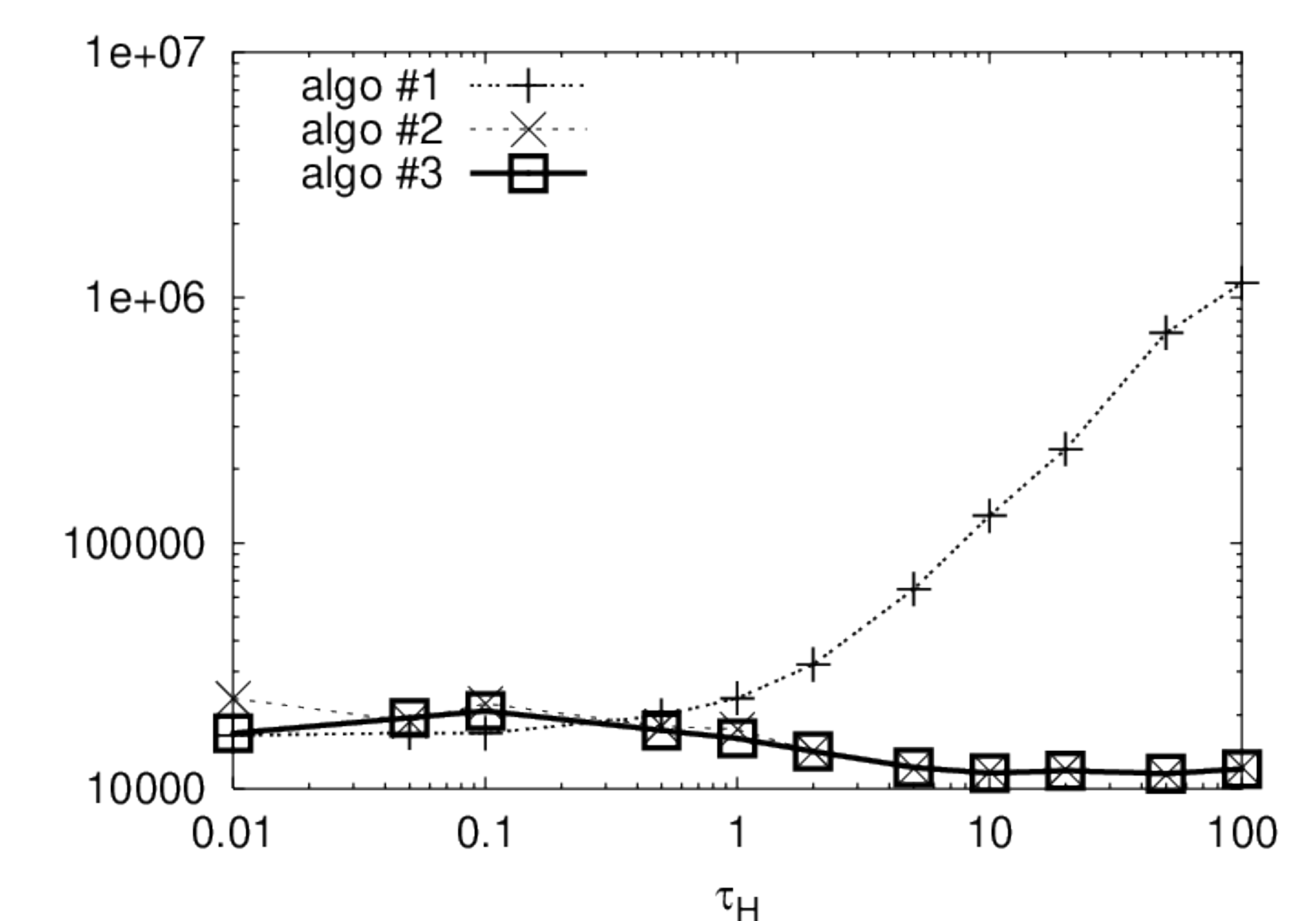,width=0.40\textwidth}}\quad
    \subfigure[$<Ns>$ for $\omega=0.50$]{\epsfig{figure=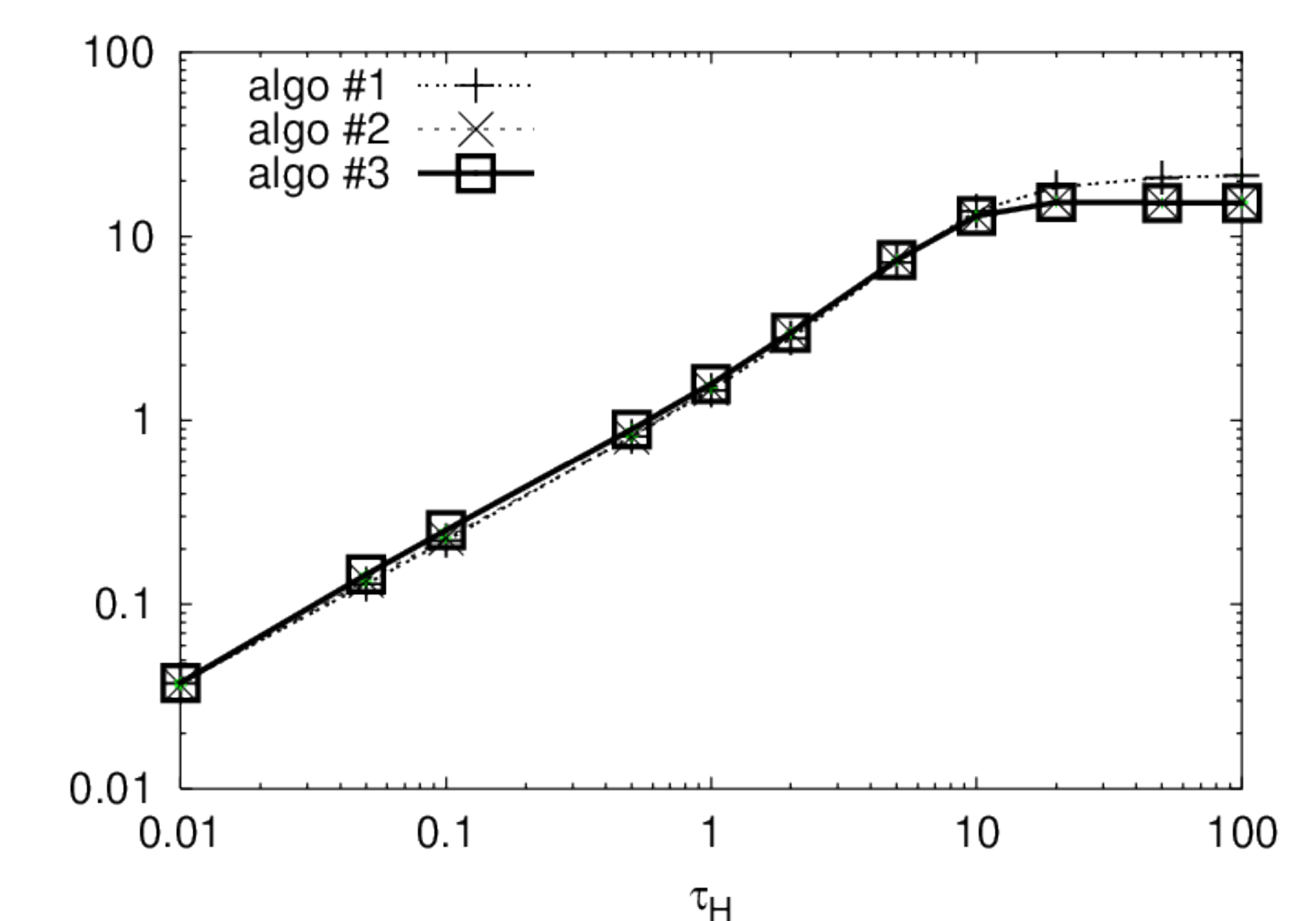,width=0.40\textwidth}}}
    \caption{Same as \fig{conv1}, except that $\omega=0.50$}
\label{fig:conv2}
\end{figure}

\begin{figure}[htbp]
\centering
\mbox{
    \subfigure[$N$ for $\omega=0.90$]{\epsfig{figure=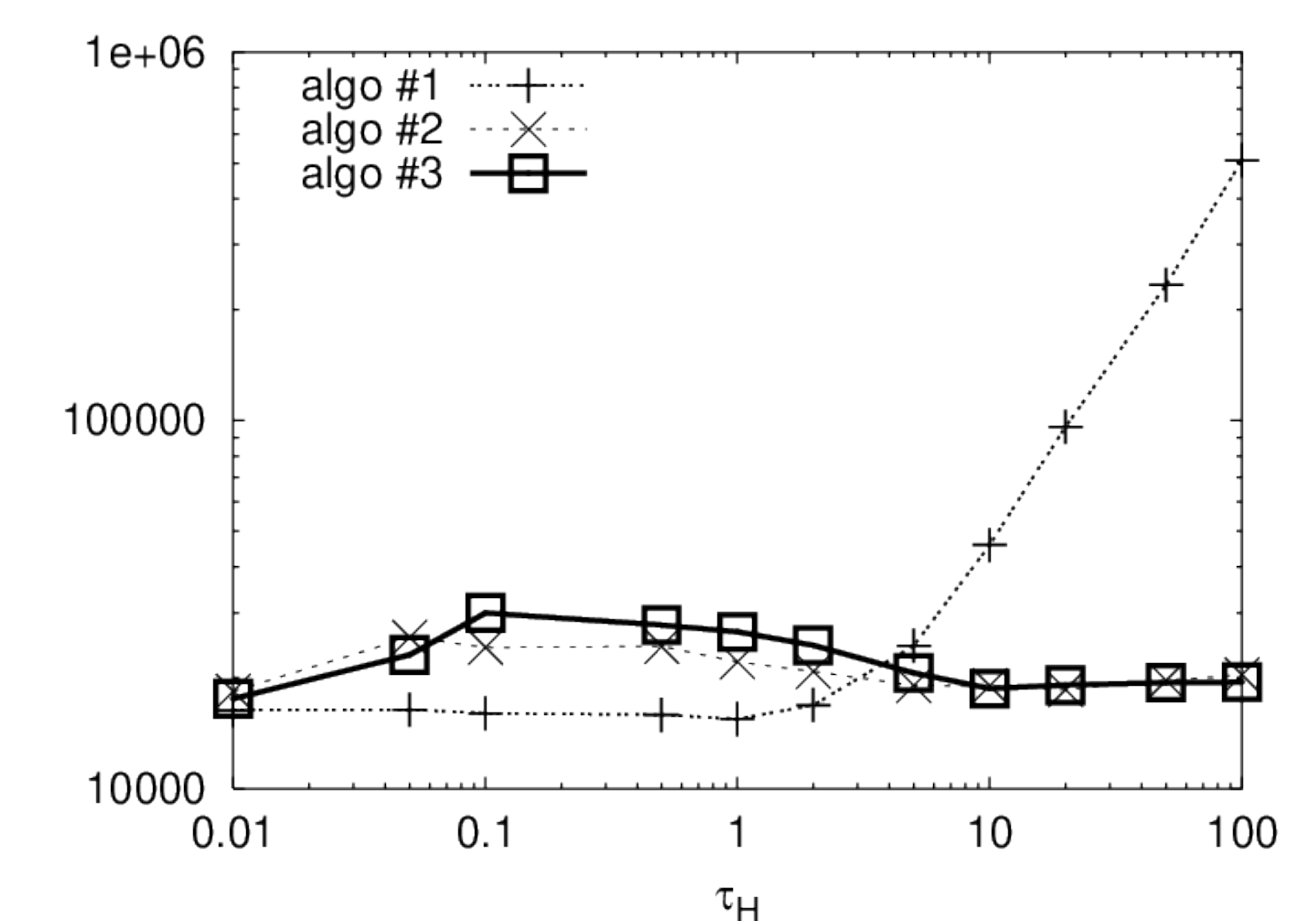,width=0.40\textwidth}}\quad
    \subfigure[$<Ns>$ for $\omega=0.90$]{\epsfig{figure=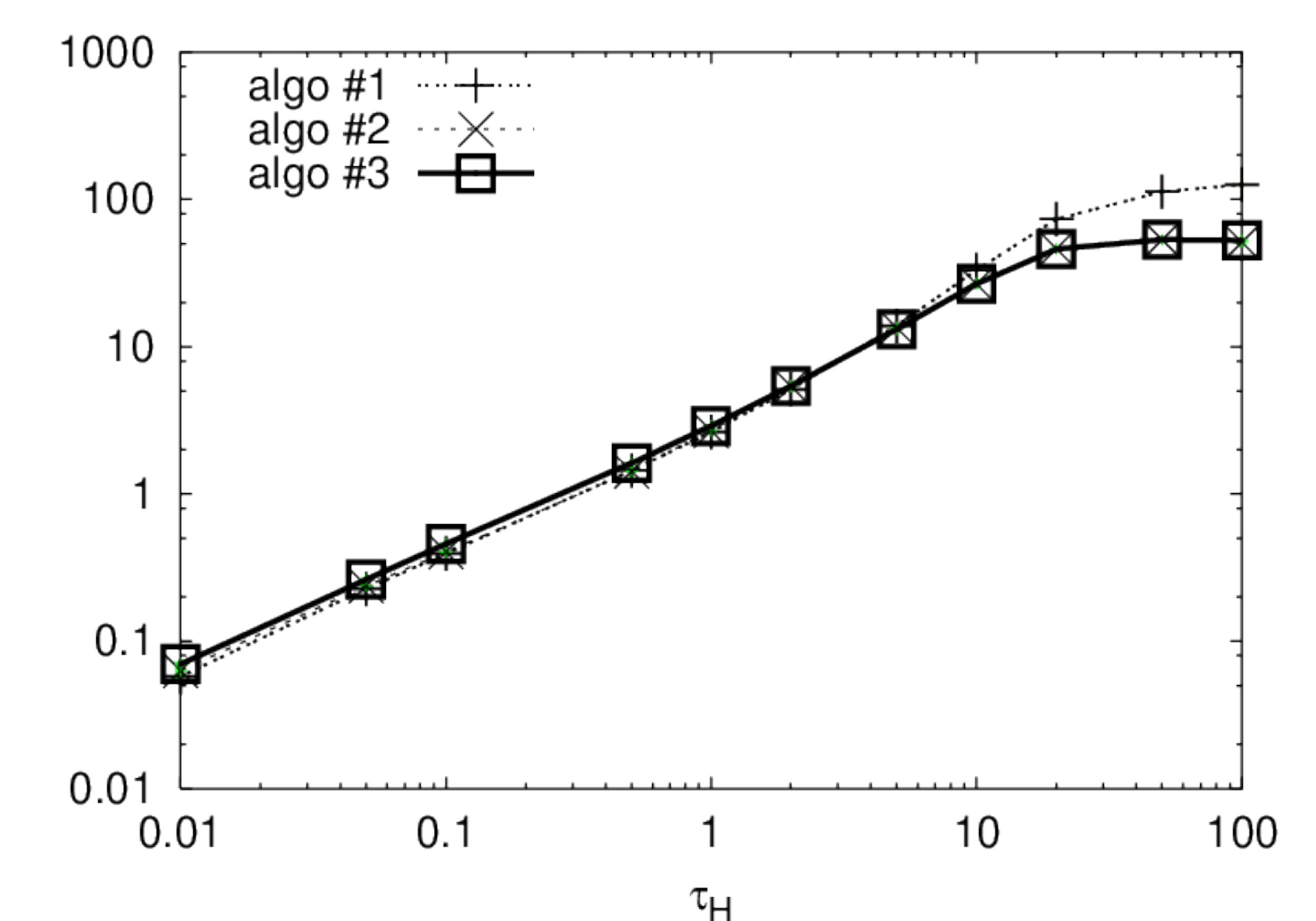,width=0.40\textwidth}}}
    \caption{Same as \fig{conv1}, except that $\omega=0.90$}
\label{fig:conv3}
\end{figure}

\begin{figure}[htbp]
\centering
\mbox{
    \subfigure[$N$ for $\omega=0.9999$]{\epsfig{figure=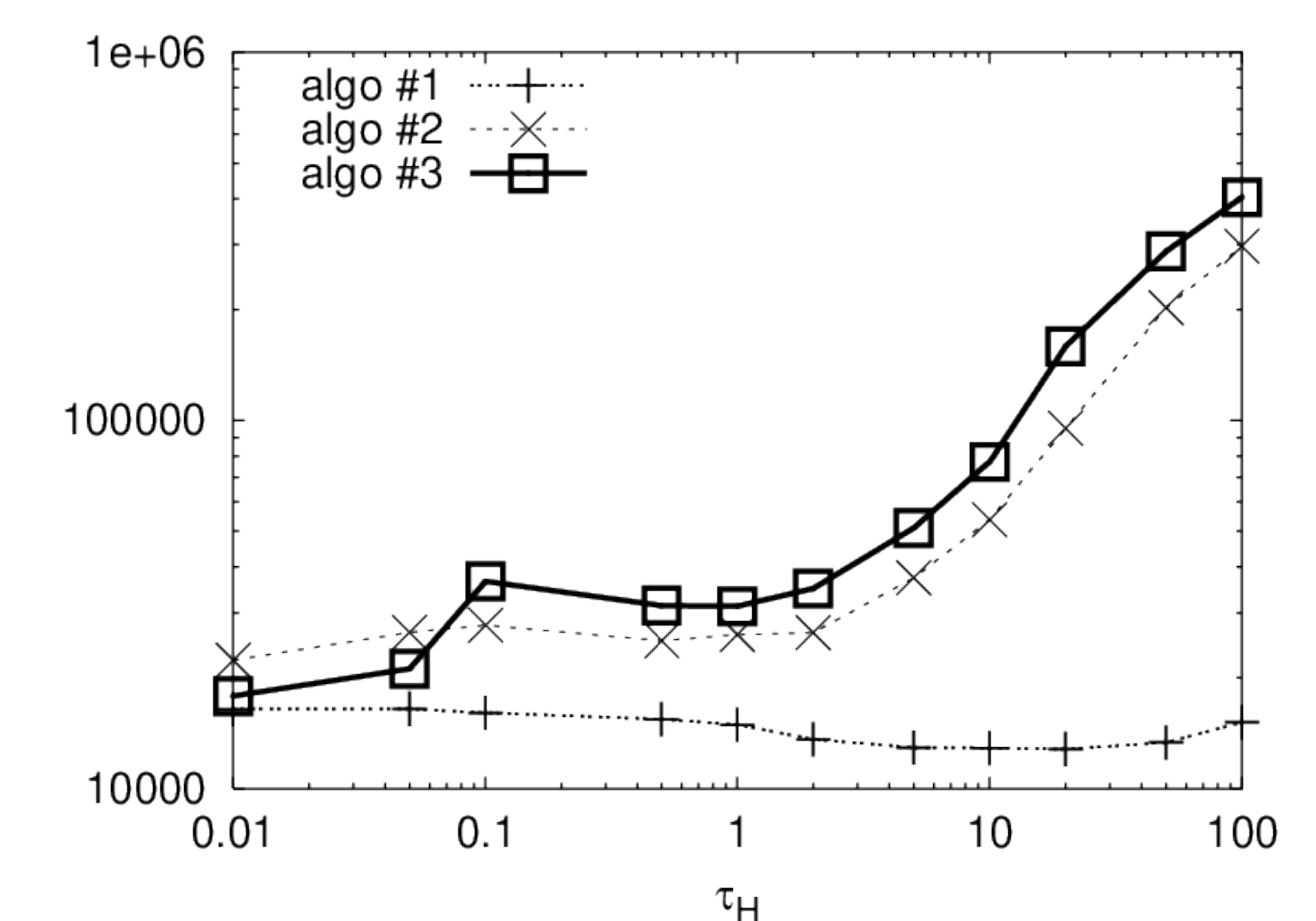,width=0.40\textwidth}}\quad
    \subfigure[$<Ns>$ for $\omega=0.9999$]{\epsfig{figure=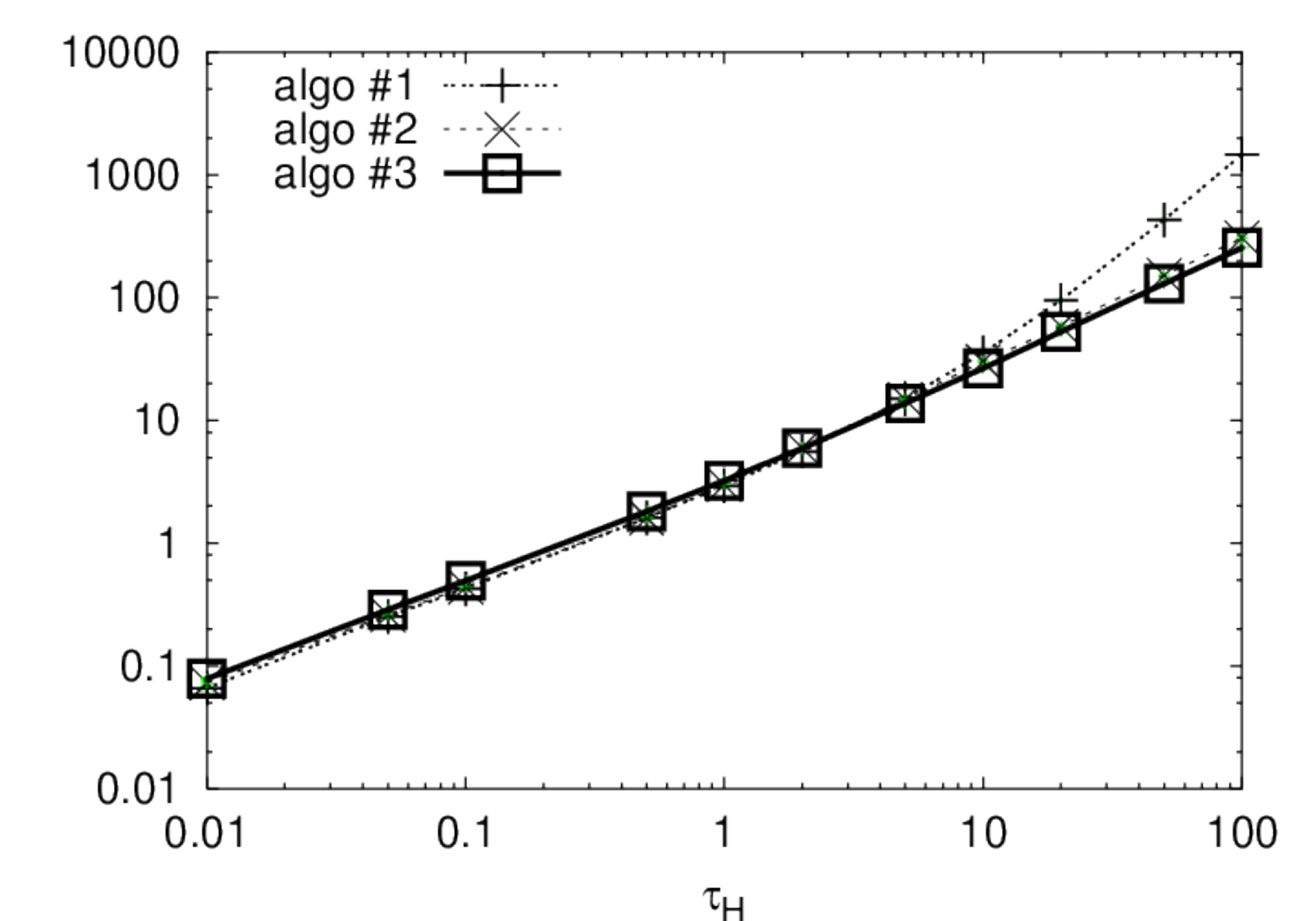,width=0.40\textwidth}}}
    \caption{Same as \fig{conv1}, except that $\omega=0.9999$}
\label{fig:conv4}
\end{figure}

\fig{conv1}a-\ref{fig:conv4}a display the number of statistical realizations $N$ needed in order to get a $1$ percent standard deviation over the slab emission value, as a function of slab total optical thickness $\tau_{H}$, for 4 different values of the single scattering albedo $\omega=\frac{k_{s}}{k_{a}+k_{s}}$. Correspondingly, \fig{conv1}b-\ref{fig:conv4}b display the mean number of scattering events $<N_{s}>$ along each sampled optical path.


In each figure, $N$ is displayed for three different Monte-Carlo algorithms~: 
\begin{itemize}
\item 1 - A standard Monte-Carlo algorithm, in which bundles are generated uniformly within the layer, with isotropic directions, and are attenuated along their multiple scattering optical paths until they leave the layer (algorithm based on an exchange formulation with a uniform law for volume sampling and an isotropic law for angular sampling, see \eq{Eij}).
\item 2 - The boundary-based net-exchange algorithm proposed in \para{part1}.
\item 3 - The same algorithm except that the angular sampling law of \cite{amaury02} is used (see \eq{dir1}), instead of that in which we attempted to account for scattering (see \eq{dir2}).
\end{itemize}

It can be seen in \fig{conv1}(a) that for small values of the single-scattering albedo ($\omega=0.01$), $N$ is stabilizing for algorithms 2 and 3 (boundary based algorithms) as the slab total optical thickness $\tau_{H}$ increases, while for algorithm 1 (standard MC algorithm), $N$ keeps increasing for large values of $\tau_{H}$. In the case of intermediate single-scattering albedoes (\fig{conv2}(a), $\omega=0.50$) and even for moderately strong single-scattering albedoes (\fig{conv3}(a), $\omega=0.90$), convergence with a $1$ percent error always requires a lower number of statistical realizations for algorithms 2 and 3.

It is no longer the case for extremely strong single-scattering albedoes (\fig{conv4}(a), $\omega=0.9999$) ; this convergence difficulty for high albedoes was explained in the previous section~: for a high value of $\omega$, the medium is optically thin for absorption, and first exchange points $P$ should be sampled uniformly within the slab. This is what the standard algorithm does, whereas most optical paths sampled by algorithms 2 and 3 (starting from the slab boundaries) are very short (because of the medium's strong scattering coefficient) thus first exchange positions $P$ are mainly sampled close to the boundaries. Altogether, in the limit of extremely high albedoes, algorithms 2 and 3 require a greater number of statistical realizations because of a non-adapted $P$ sampling law.

\begin{figure}[htbp]
\centering
\mbox{
    \subfigure[Complete $\omega$ range]{\epsfig{figure=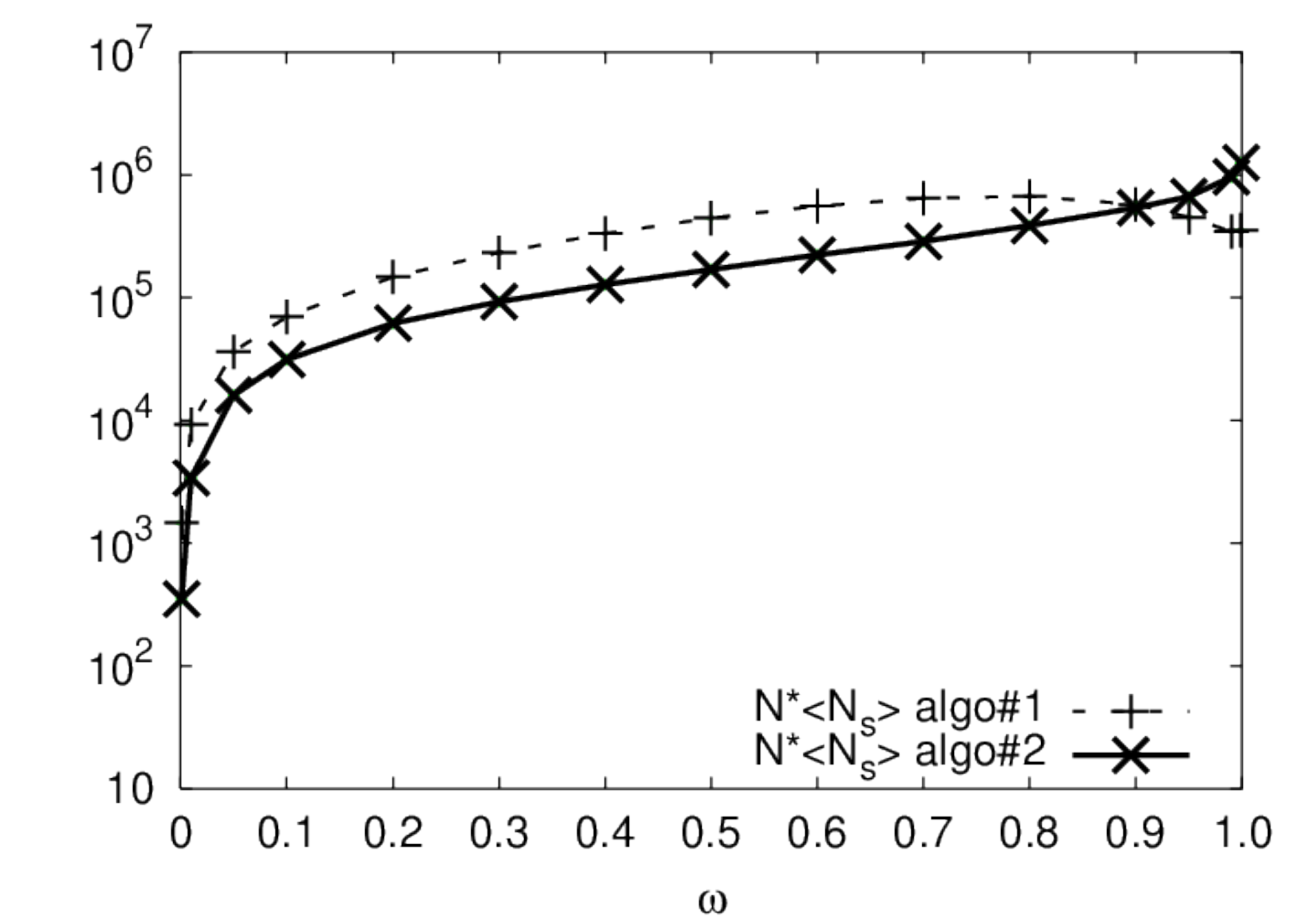,width=0.40\textwidth}}\quad
    \subfigure[zoom over 0.90-0.92 range]{\epsfig{figure=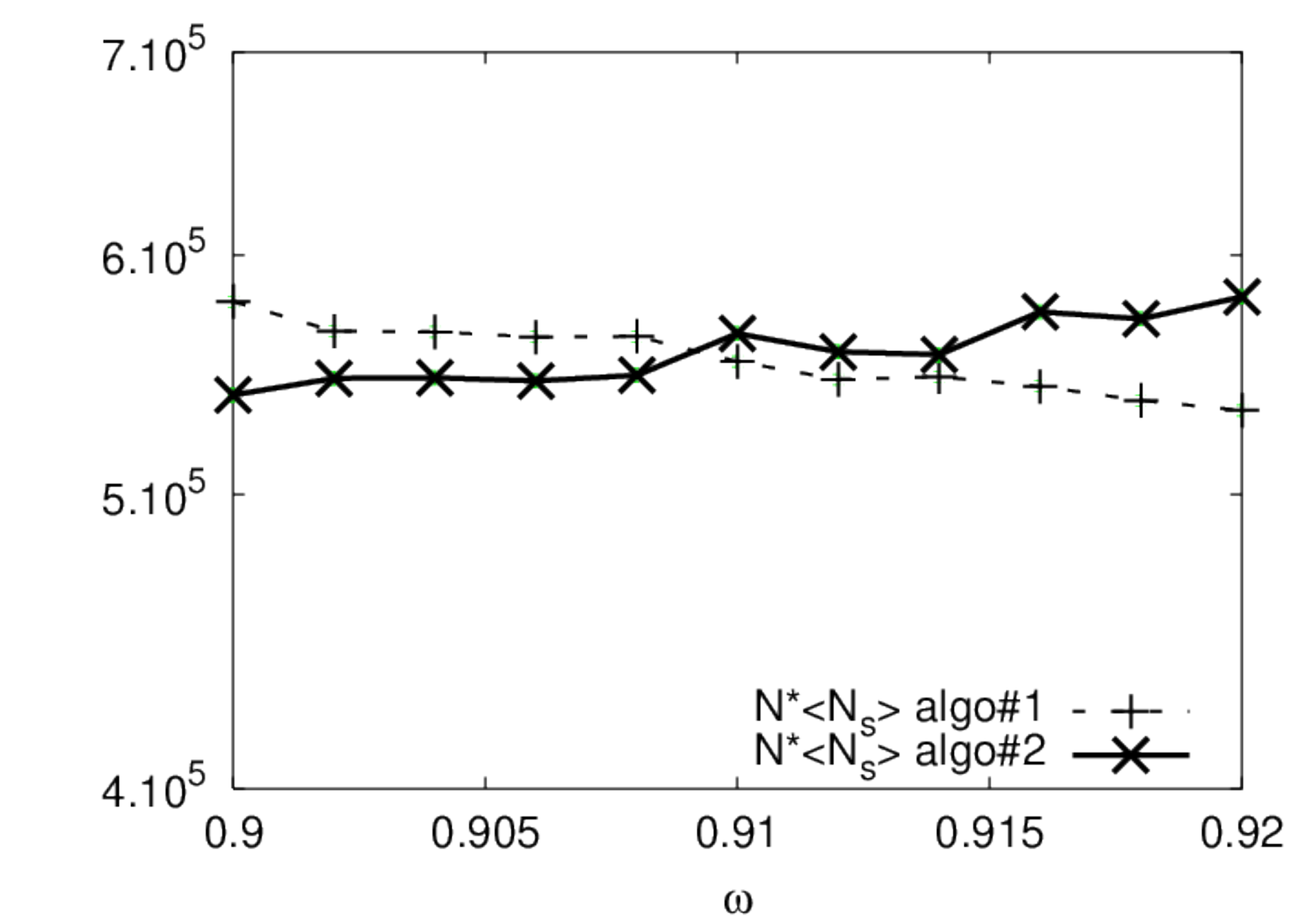,width=0.40\textwidth}}}
    \caption{$N<N_{s}>$ for $\tau_{H}=10$}
\label{fig:speed10}
\end{figure}

\begin{figure}[htbp]
\centering
\mbox{
    \subfigure[Complete $\omega$ range]{\epsfig{figure=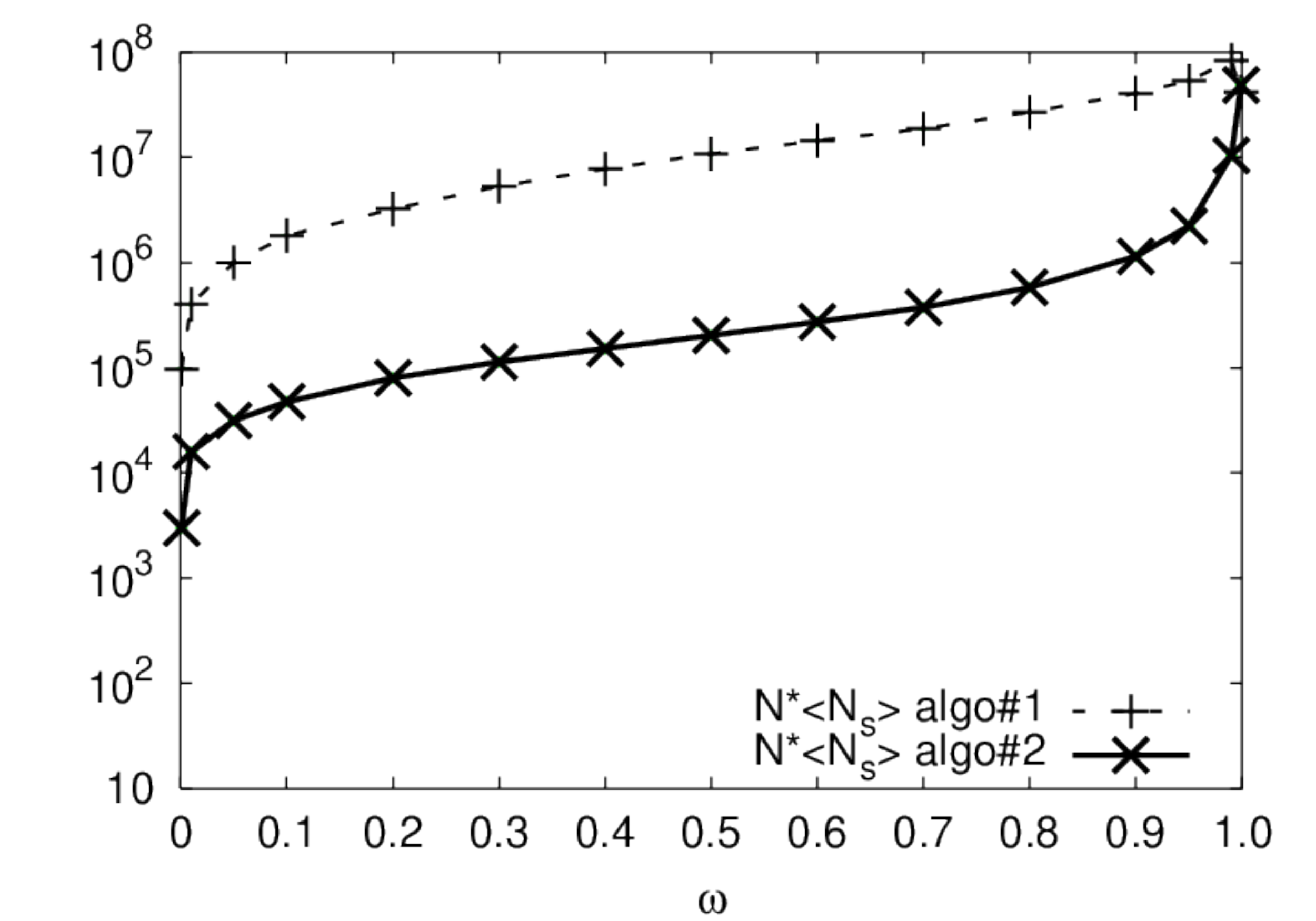,width=0.40\textwidth}}\quad
    \subfigure[zoom over 0.98-1 range]{\epsfig{figure=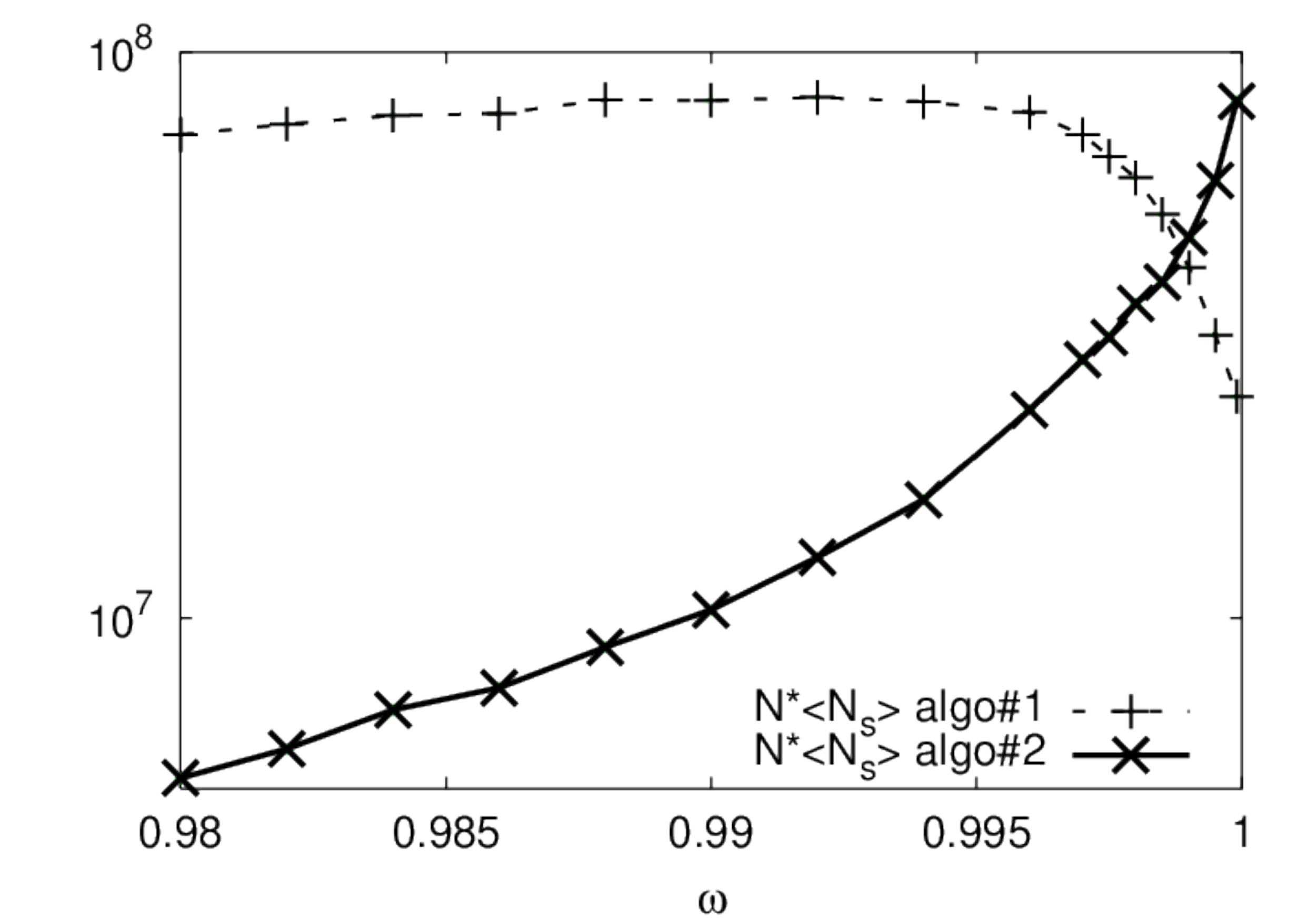,width=0.40\textwidth}}}
    \caption{$N<N_{s}>$ for $\tau_{H}=100$}
\label{fig:speed100}
\end{figure}

However, the numerical cost of the algorithm is not directly the number of required statistical realizations $N$, but the product $N <N_{s}>$ where $<N_{s}>$ is the average number of scattering events. Concerning $<N_{s}>$, \fig{conv1}(b) - \ref{fig:conv4}(b) illustrate that~:
\begin{itemize}
\item For low values of $\tau_{H}$ and $\omega$, the mean number of scattering events $<N_{s}>$ required for each statistical realization is of the same order of magnitude for all three algorithms.
\item In the special case of both high $\tau_{H}$ and high $\omega$, $<N_{s}>$ can be about $10$ times greater for the standard algorithm than for algorithms 2 and 3.
\end{itemize}
This may be explained, making the assumption that \mbox{$<N_{s}> \approx \frac{<L>}{\lambda_{s}} = <L>k_{s}$} with $<L>$ the average path length and $\lambda_{s}=\frac{1}{k_{s}}$ the scattering mean free path. For algorithms 2 and 3, it has been shown (see. \cite{Blanco01}) that $<L>$ is independent of scattering properties : $<L>=2H$. For algorithm 1, it can be easily shown that $<L>$ is proportional to $H\tau_{s}$.\footnote{This property may be derived directly from Markov theory with absorbing states \cite{Feller} in a one-dimensional case, with constant free path length (problem well known as the ``Gambler's ruin problem''). Extension to exponentially distributed free path length is tedious but is accessible without any specific mathematical difficulty. To our knowledge, extension to three dimensions is not available, but it may easily be observed experimentally that the proportionality property remains valid, at least for qualitative reasonings such as those made in the present text.} At high values of $\omega$, this finally gives $<N_{s}>\thicksim\tau_{H}^{2}$ for algorithm 1 and $<N_{s}> \approx 2\tau_{H}$ for algorithms 2 and 3.

These two competing effects combine at high albedo and results are shown in \fig{speed10} and \fig{speed100}. These figures display the product $N<N_{s}>$ for $\tau_{H}=10$ and $\tau_{H}=100$, as a function of the single scattering albedo $\omega$. It appears that the two effects previously emphasized for high albedo ($N$ lower for algorithm 1 than for algorithms 2 and 3, and $<N_{s}>$ greater for algorithm 1 than for algorithms 2 and 3) result in the fact that algorithm 2 remains faster than algorithm 1 up to relatively high values of $\omega$, and becomes slower above a critical value of $\omega$. The value $\omega_{c}$ at which both algorithms converge at the same speed depend on $\tau_{H}$, $\omega_{c}$ increasing as $\tau_{H}$ increases ($\omega_{c} \approx 0.91$ for $\tau_{H}=10$ and $\omega_{c} \approx 0.998$ for $\tau_{H}=100$).
\section{Convergence illustration~: radiative flux divergence within a non-isothermal slab}
\label{para:part3}

\begin{figure}[htbp]
\centering
\mbox{
    \subfigure[$<div(q_{r})>/\pi \Delta B$ for $\omega=0.01$]{\epsfig{figure=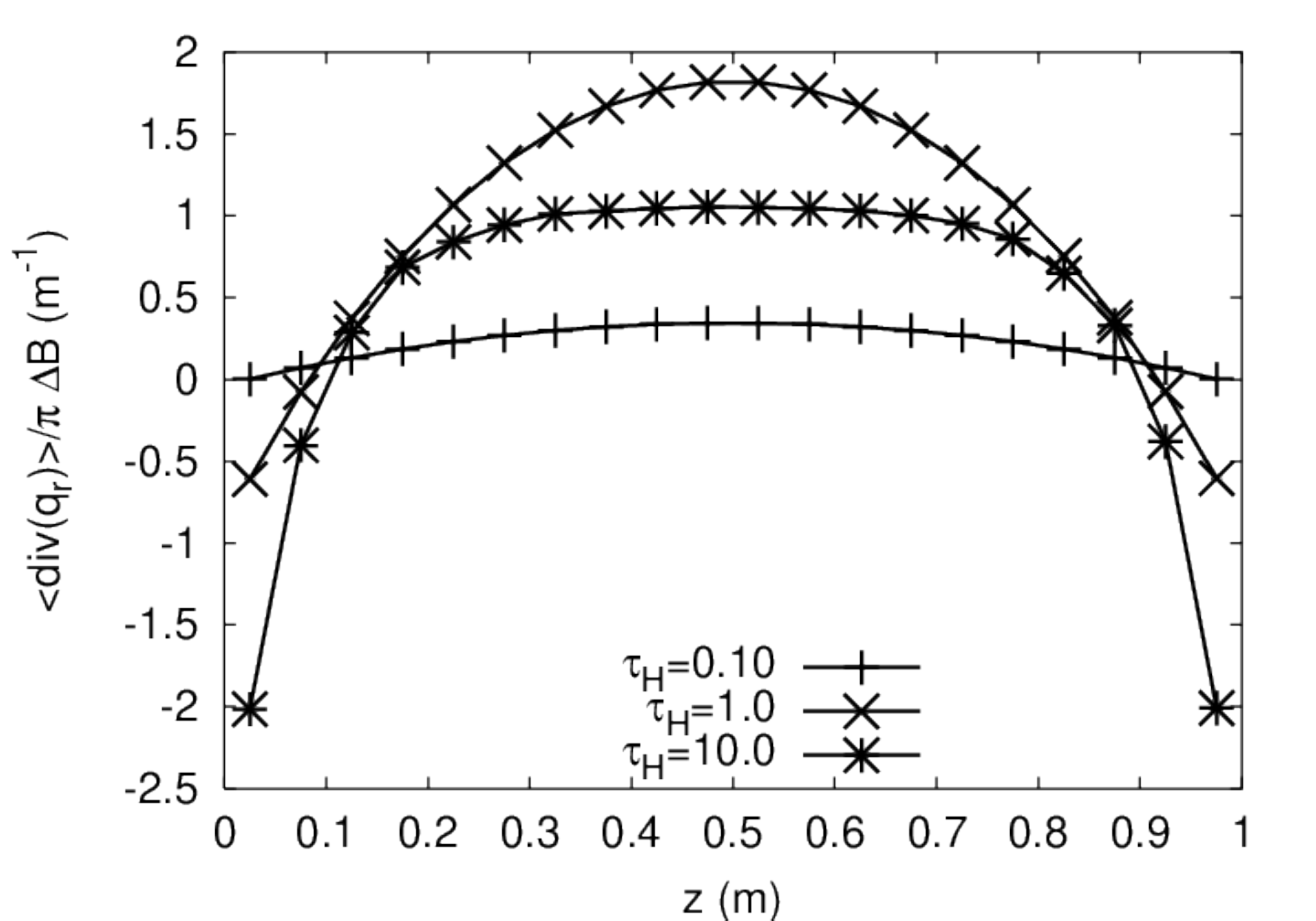,width=0.40\textwidth}}\quad
    \subfigure[statistical error]{\epsfig{figure=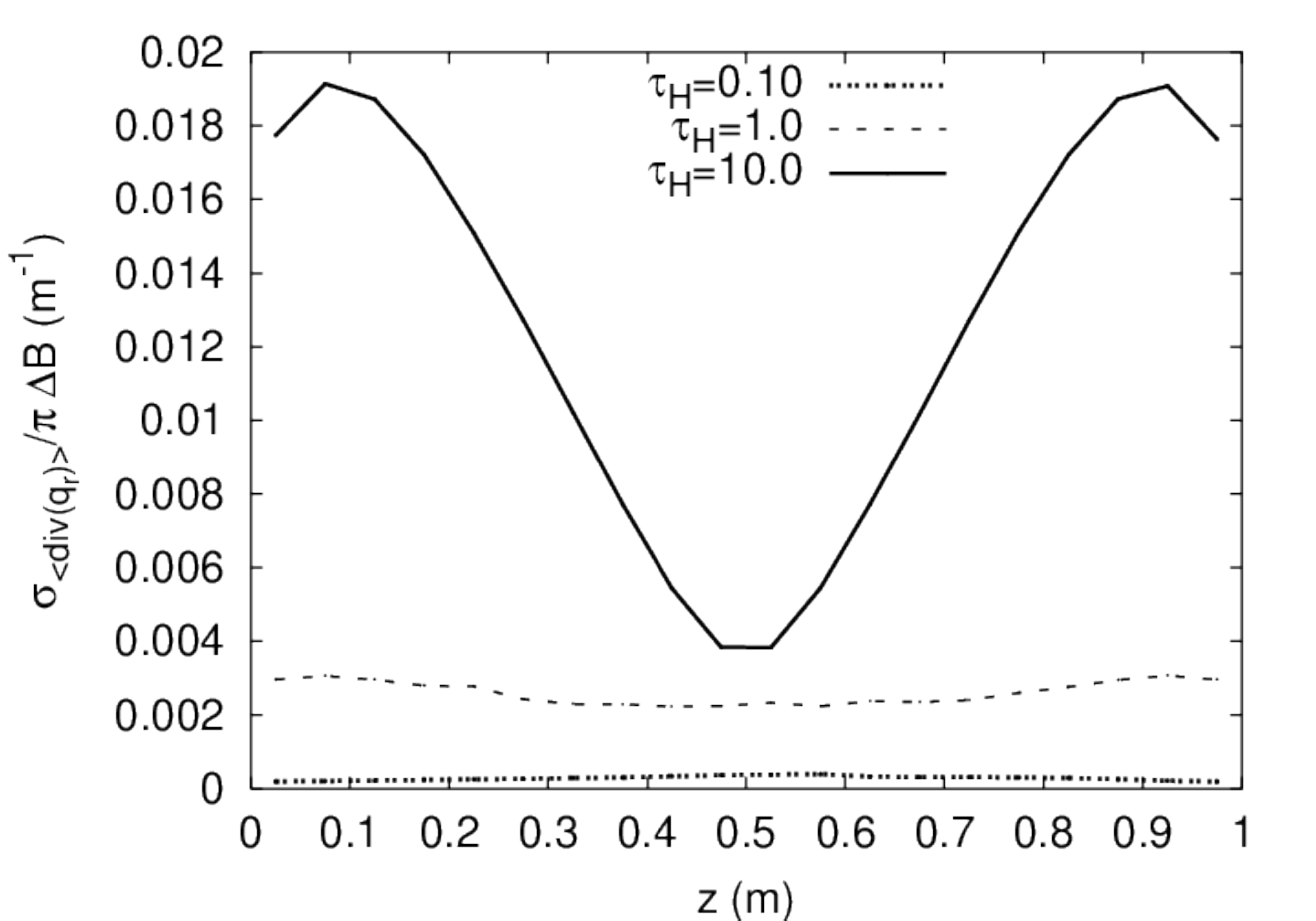,width=0.40\textwidth}}}
\mbox{
    \subfigure[$<div(q_{r})>$ layer 3 for $\omega=0.01$]{\epsfig{figure=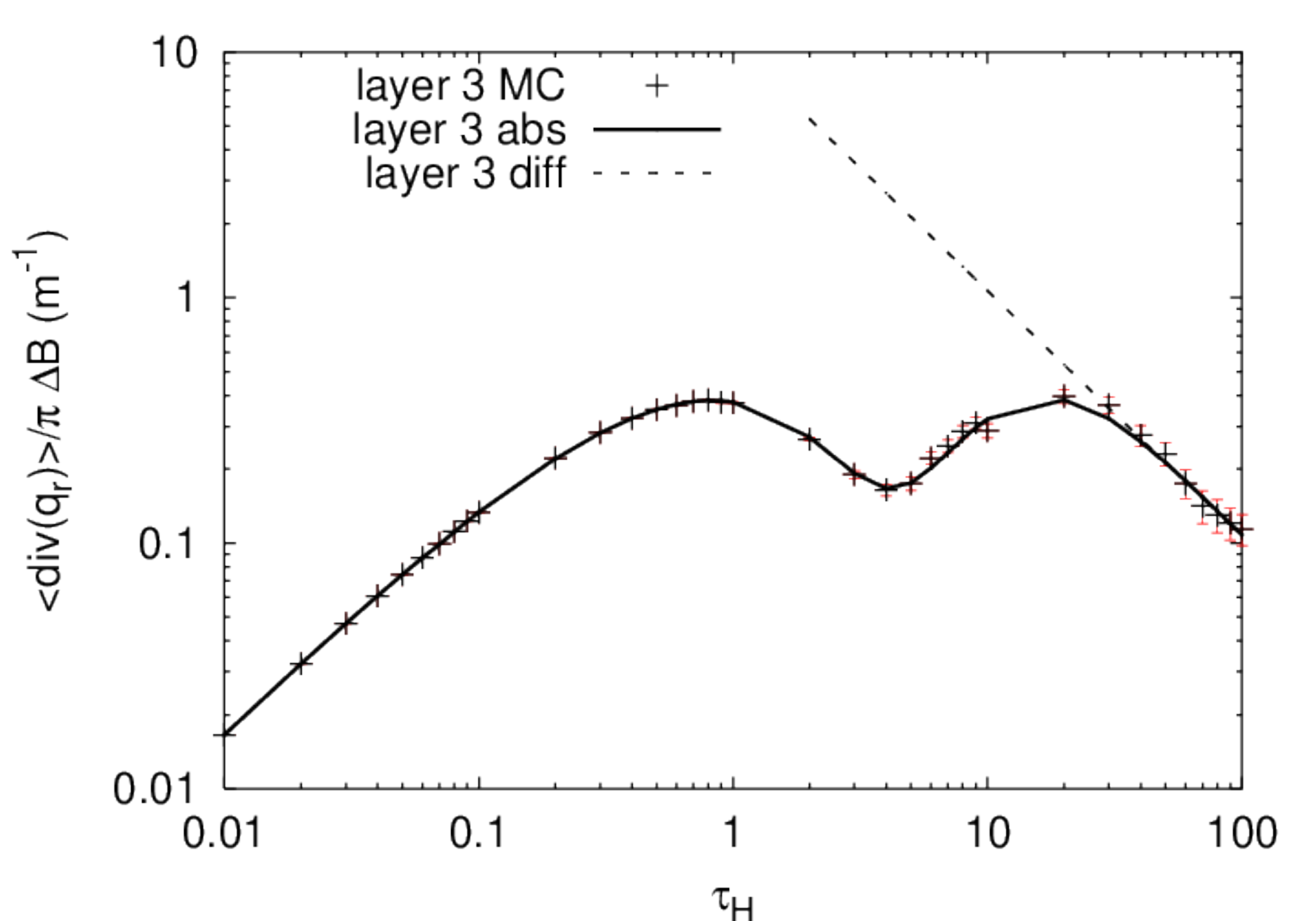,width=0.40\textwidth}}\quad
    \subfigure[$<div(q_{r})>$ layer 10 for $\omega=0.01$]{\epsfig{figure=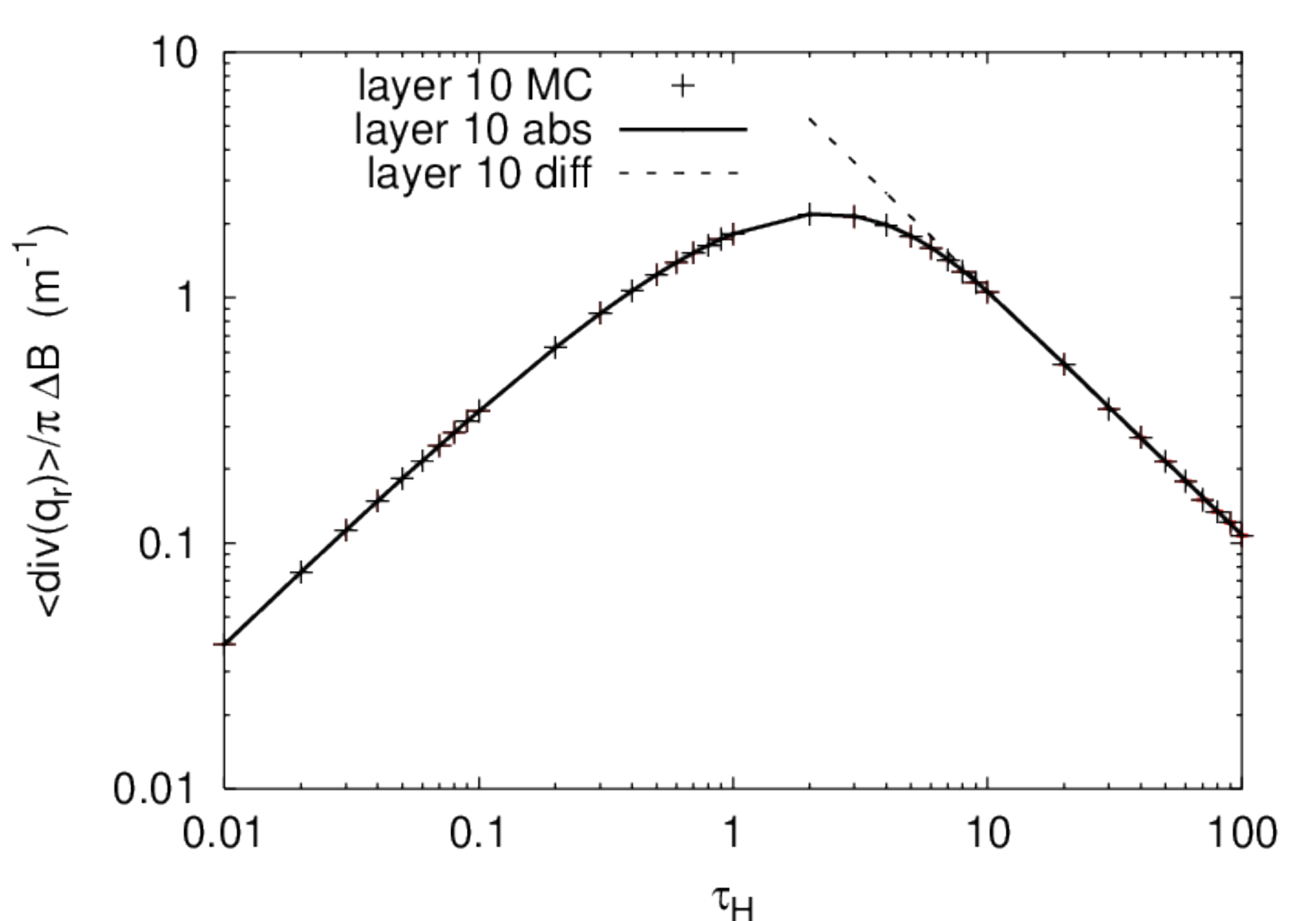,width=0.40\textwidth}}}
\mbox{
    \subfigure[percent error layer 3]{\epsfig{figure=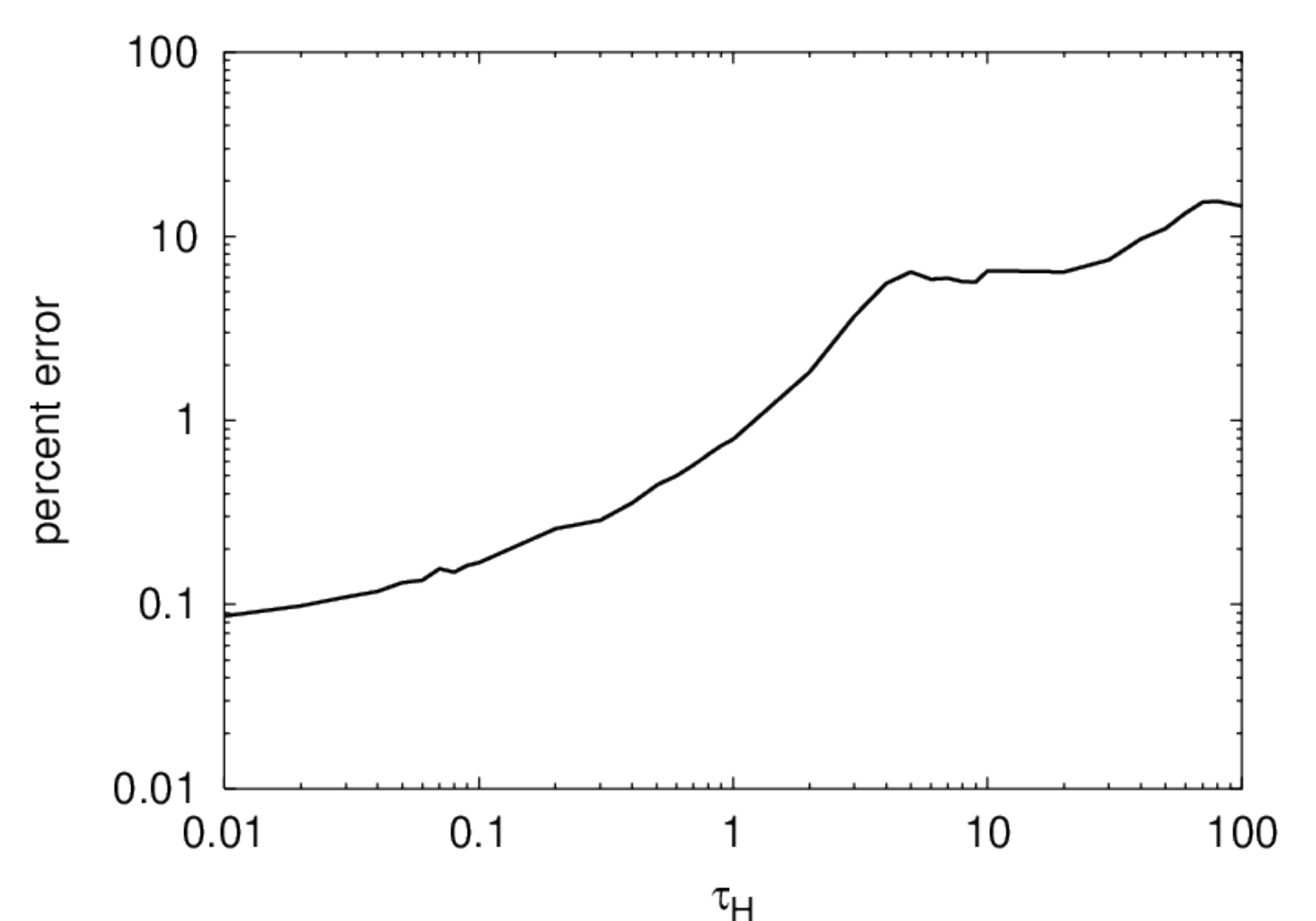,width=0.40\textwidth}}\quad
    \subfigure[percent error layer 10]{\epsfig{figure=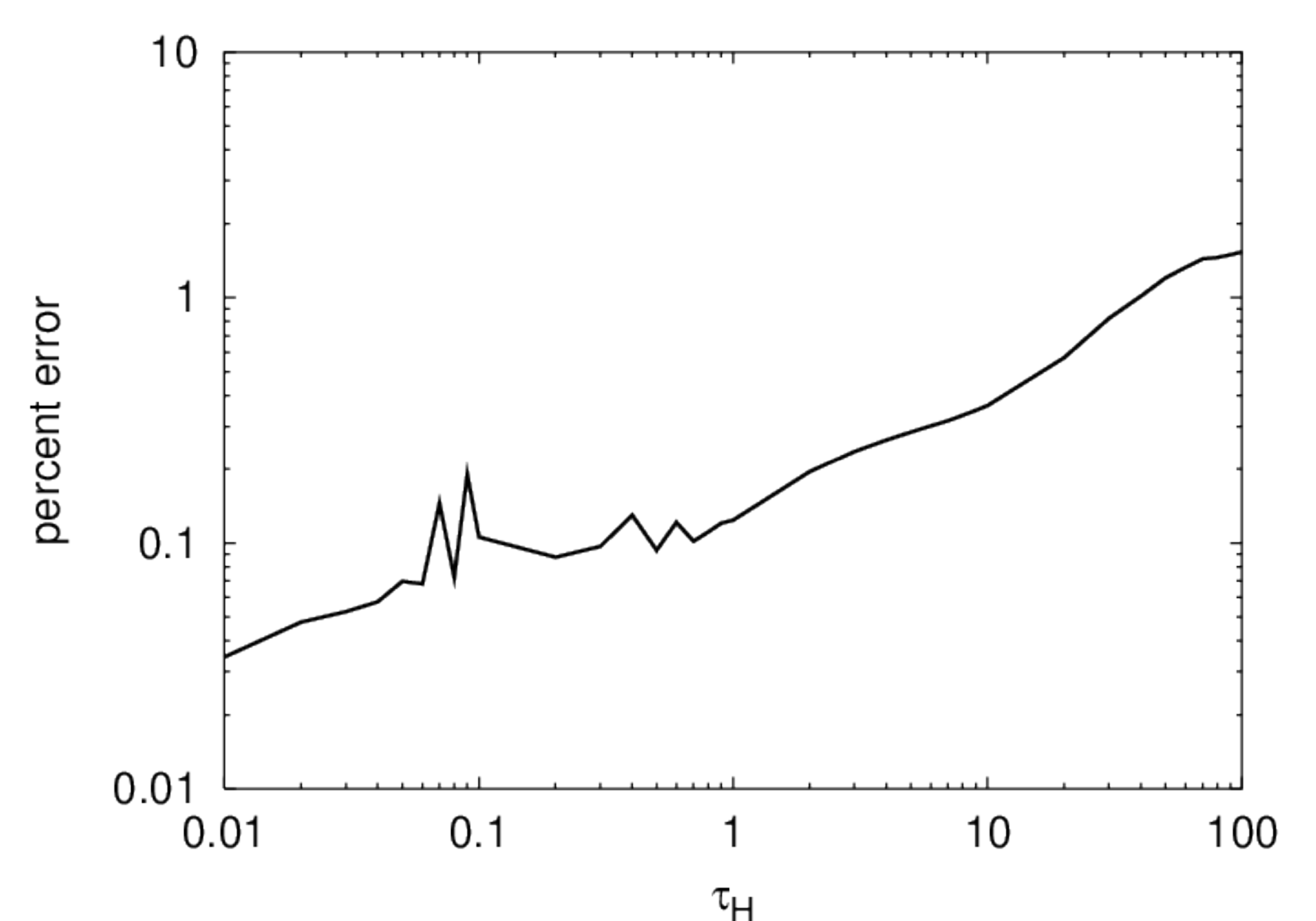,width=0.40\textwidth}}} 
\caption{Average value of the radiative flux divergence within each of the 20 layers using $N=10000$ statistical realizations per layer. The slab width is $H=1m$, scattering is isotropic and the single scattering albedo is $\omega=0.01$. (a): radiative flux divergence profile for three values of the slab total optical thickness $\tau_{H}$; (b): standard deviations corresponding to (a); (c): radiative flux divergence average in layer 3 as a function of $\tau_{H}$; (d): radiative flux divergence average in layer 10 as a function of $\tau_{H}$; (e): standard deviations corresponding to (c); (f): standard deviations corresponding to (d).}
\label{fig:para1}
\end{figure}

\begin{figure}[htbp]
\centering
\mbox{
    \subfigure[]{\epsfig{figure=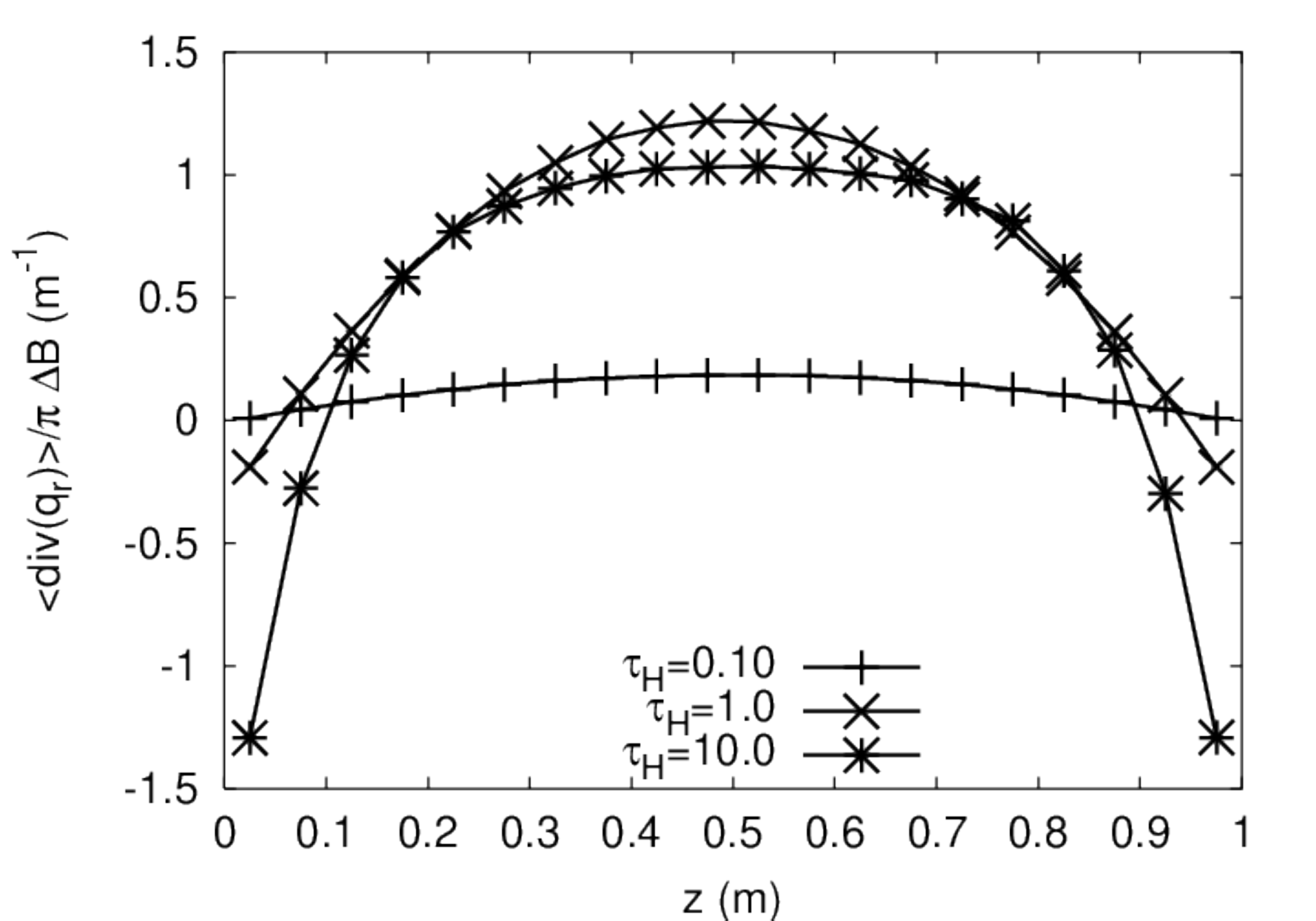,width=0.40\textwidth}}\quad
    \subfigure[]{\epsfig{figure=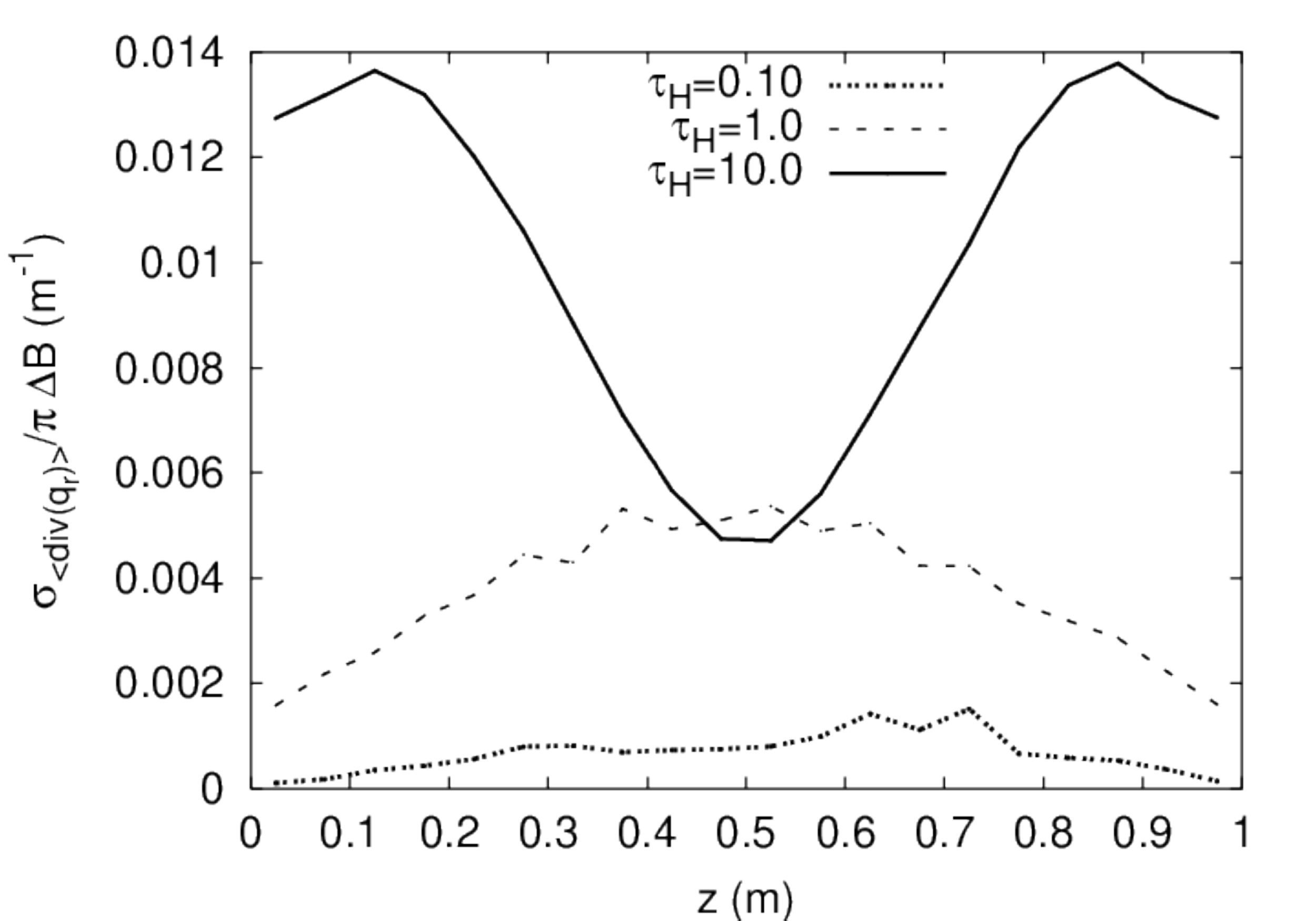,width=0.40\textwidth}}}
\mbox{
    \subfigure[Layer 3]{\epsfig{figure=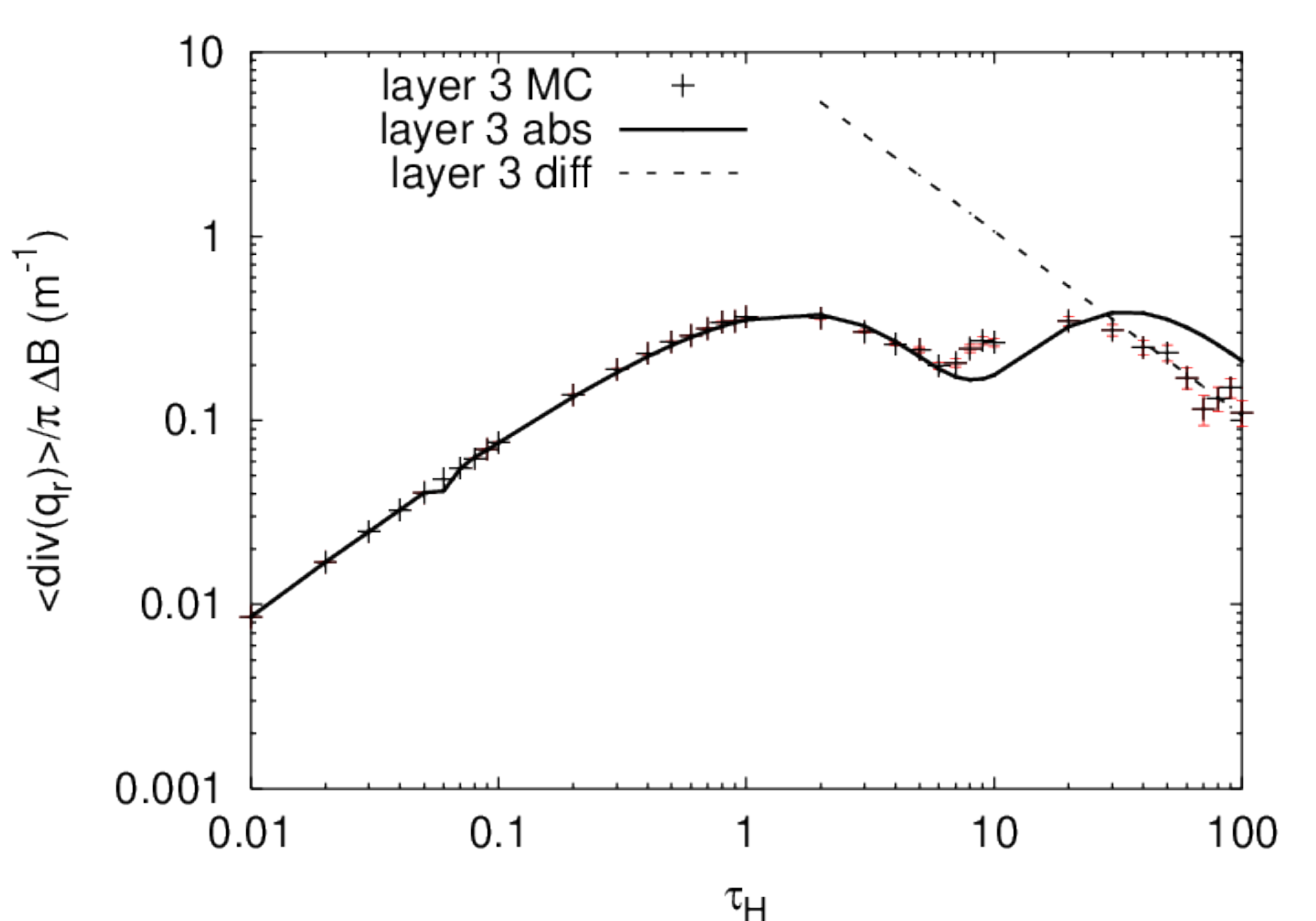,width=0.40\textwidth}}\quad
    \subfigure[Layer 10]{\epsfig{figure=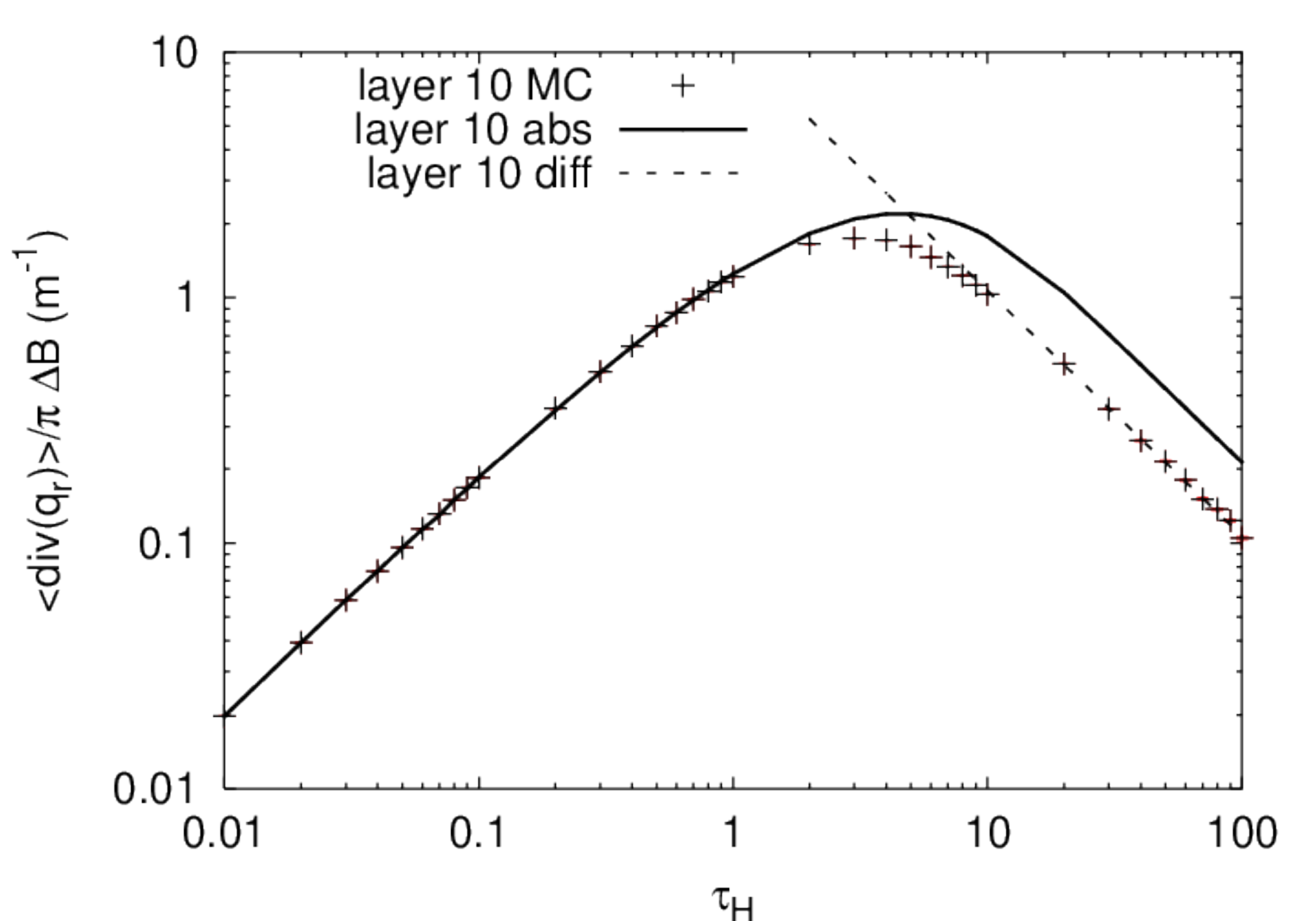,width=0.40\textwidth}}}
\mbox{
    \subfigure[Layer 3]{\epsfig{figure=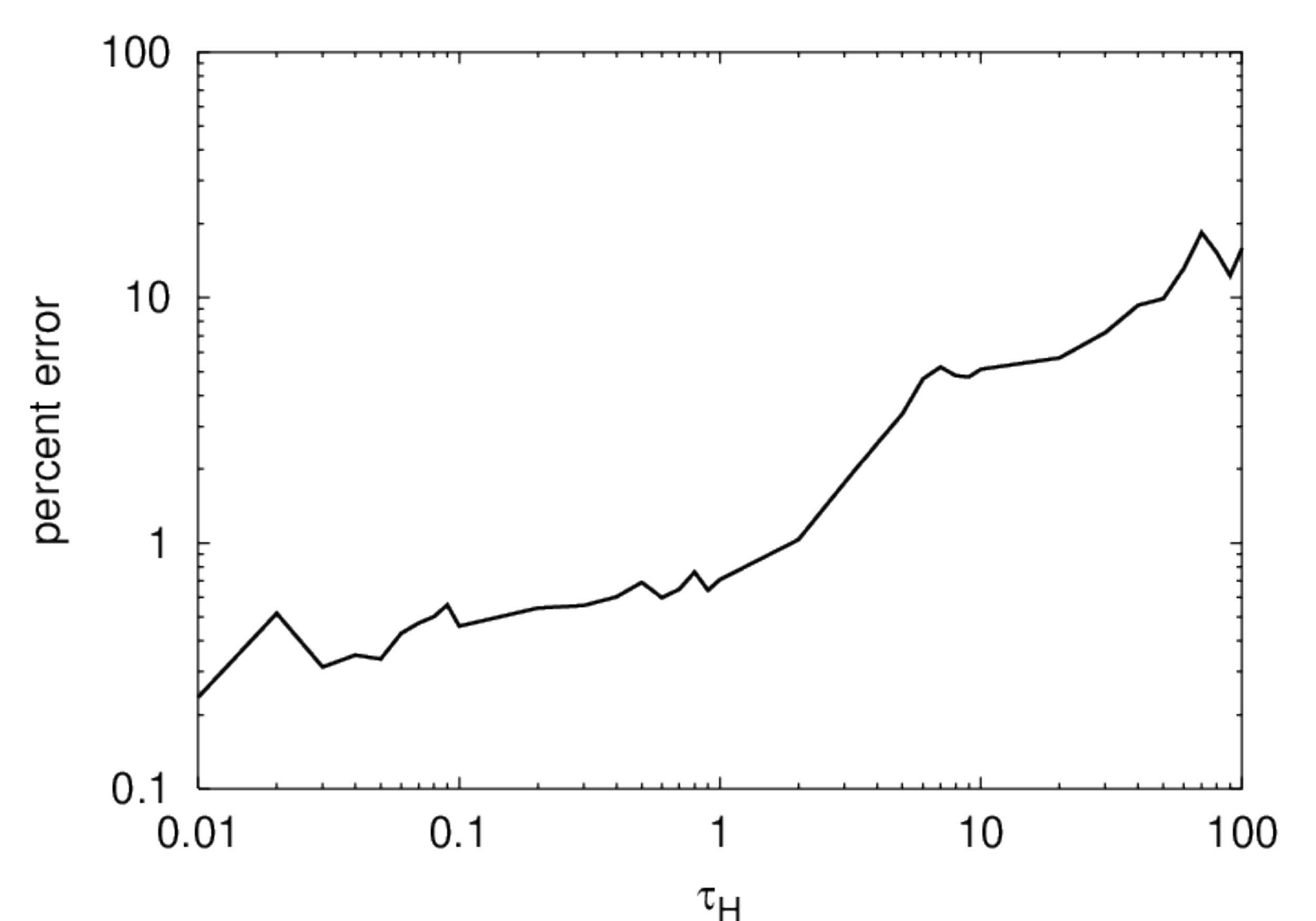,width=0.40\textwidth}}\quad
    \subfigure[Layer 10]{\epsfig{figure=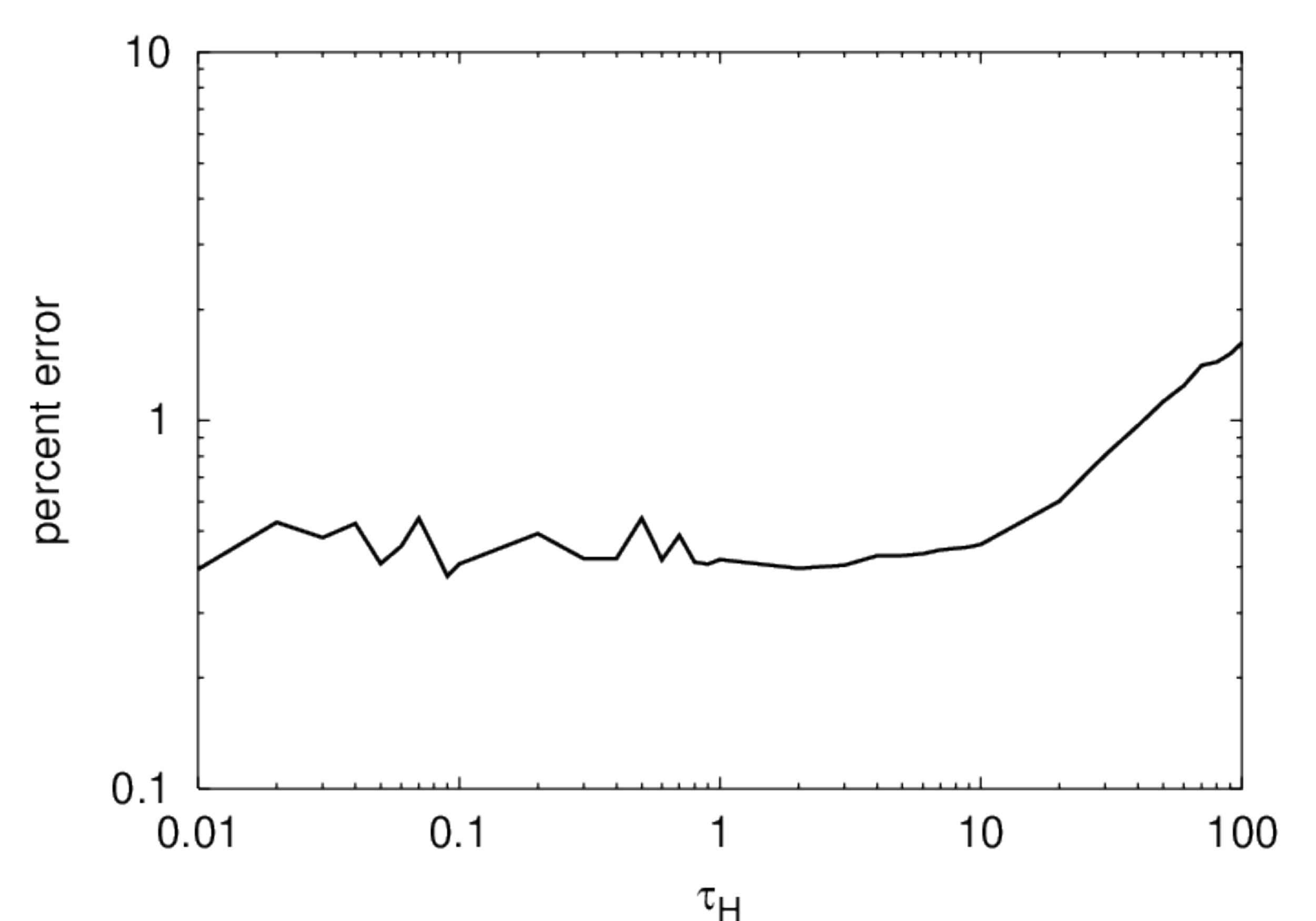,width=0.40\textwidth}}}
\caption{Same as \fig{para1} with $\omega=0.5$}
\label{fig:para2}
\end{figure}

\begin{figure}[htbp]
\centering
\mbox{
    \subfigure[]{\epsfig{figure=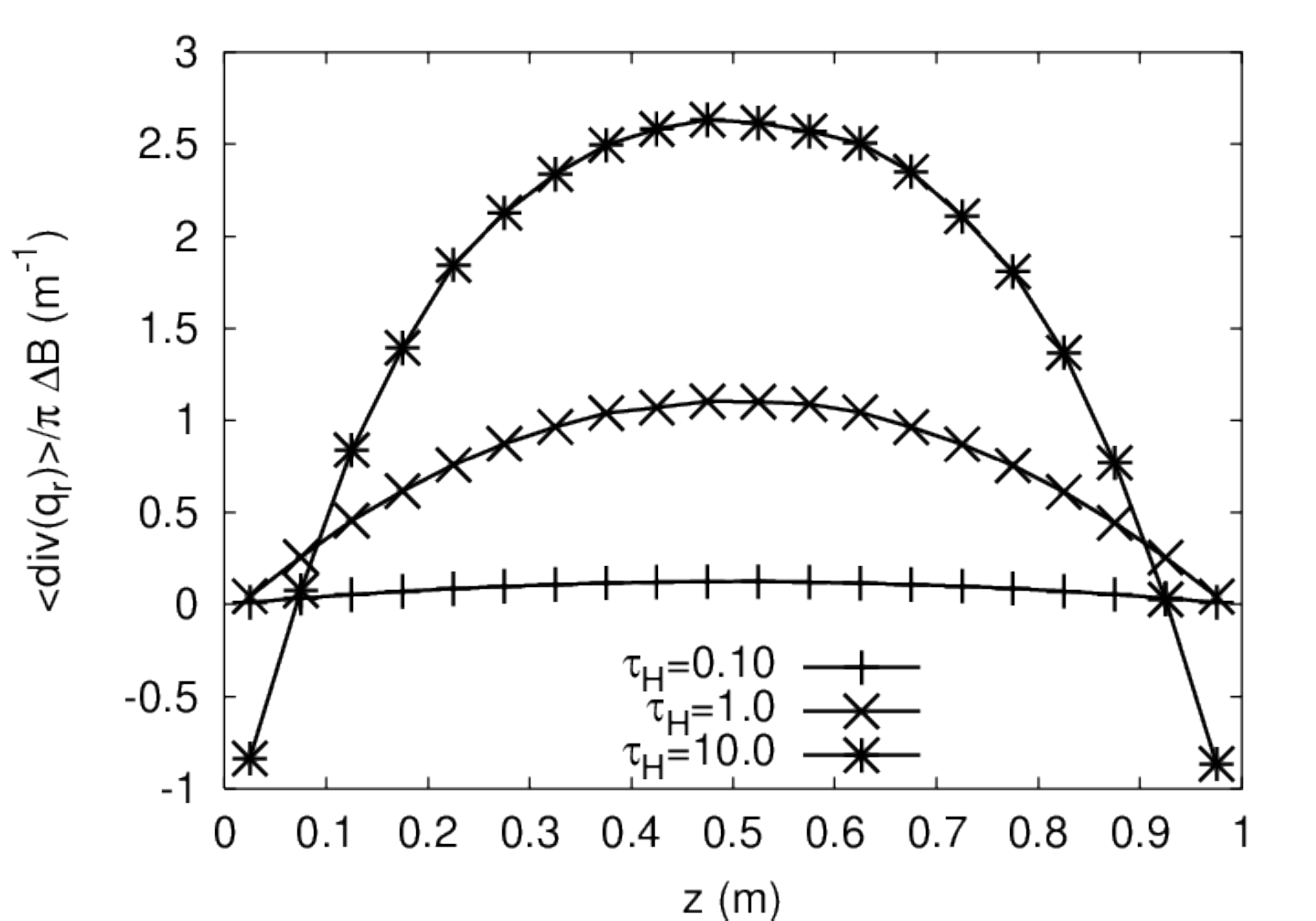,width=0.40\textwidth}}\quad
    \subfigure[]{\epsfig{figure=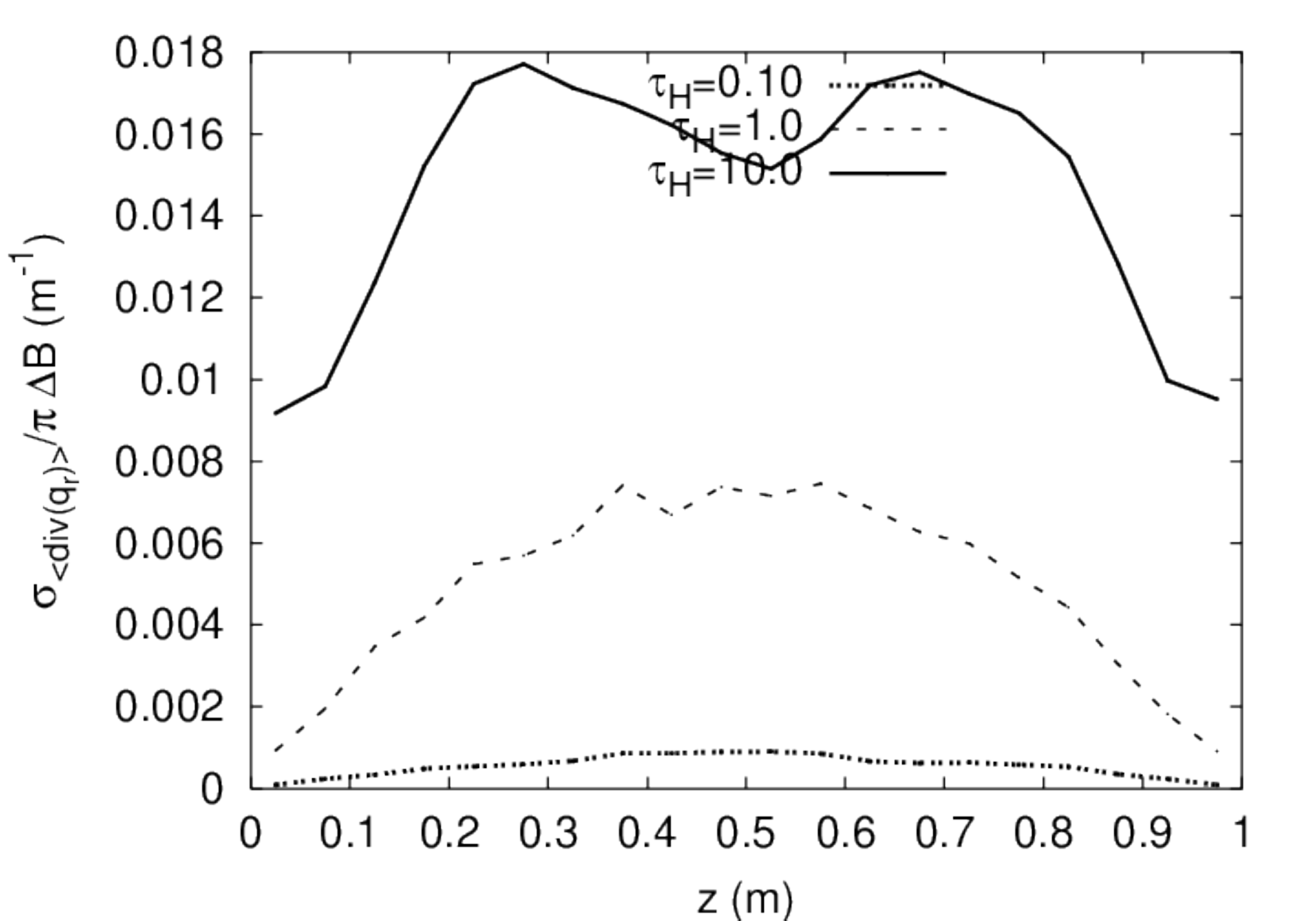,width=0.40\textwidth}}}
\mbox{
    \subfigure[Layer 3]{\epsfig{figure=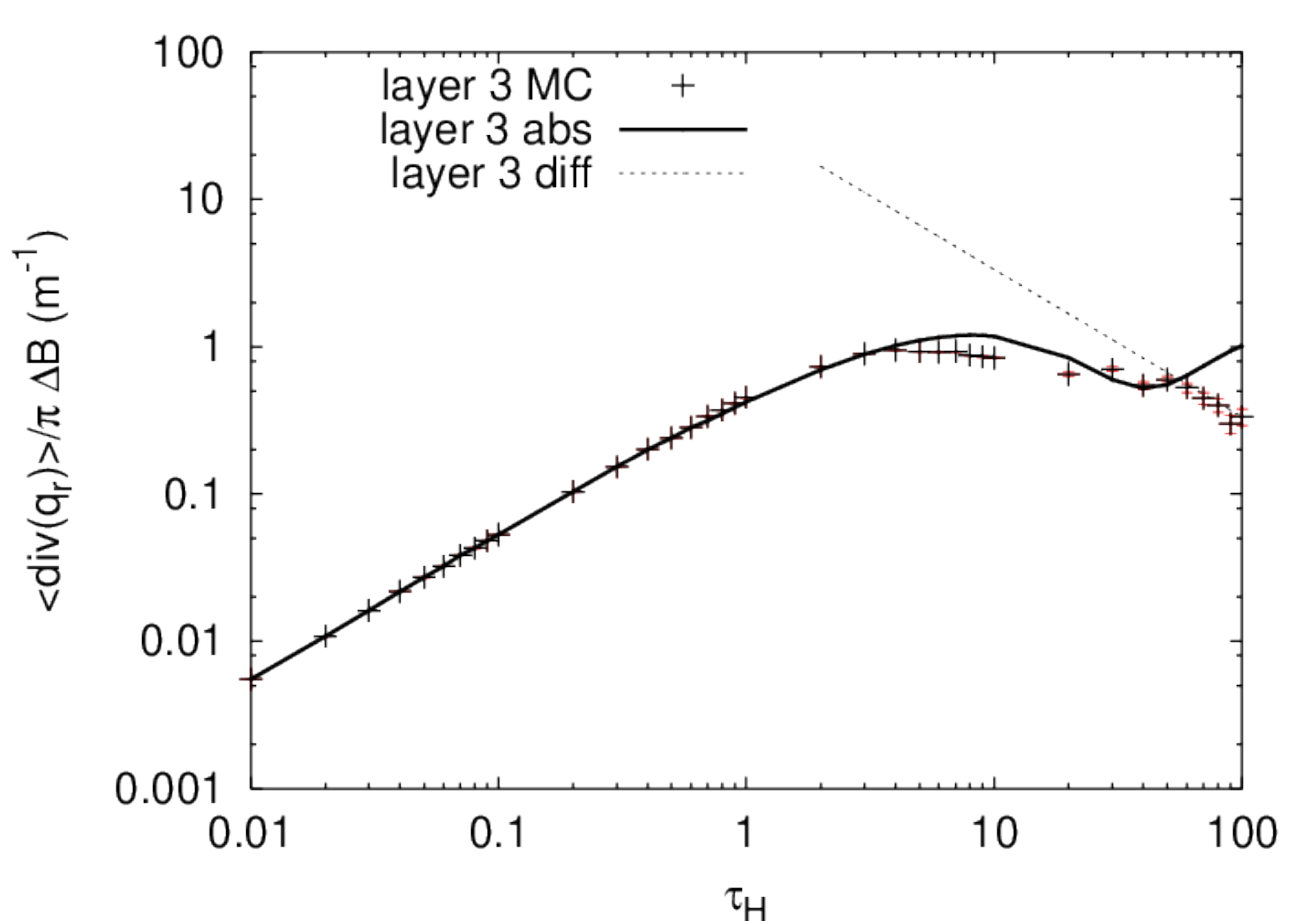,width=0.40\textwidth}}\quad
    \subfigure[Layer 10]{\epsfig{figure=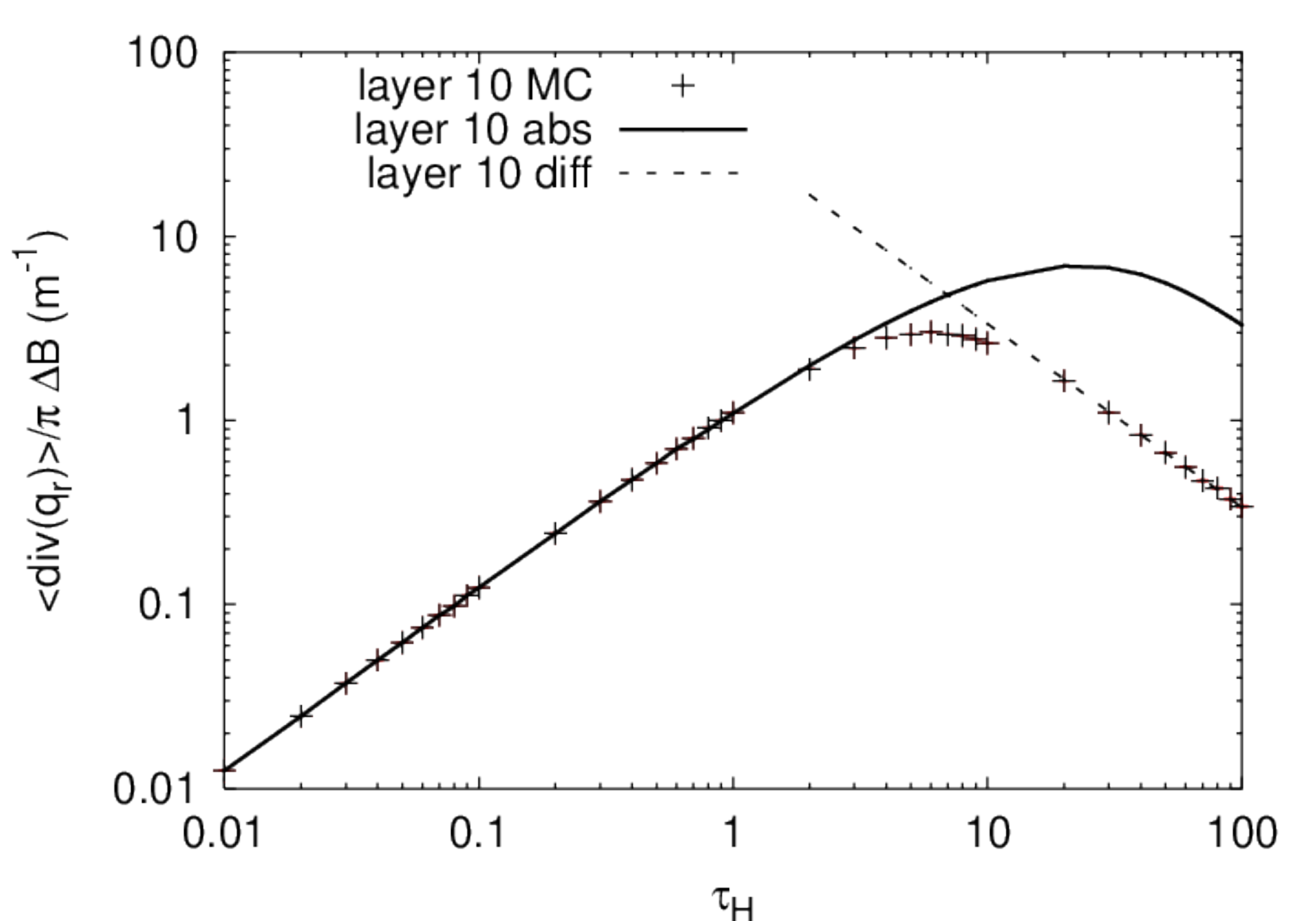,width=0.40\textwidth}}}
\mbox{
    \subfigure[Layer 3]{\epsfig{figure=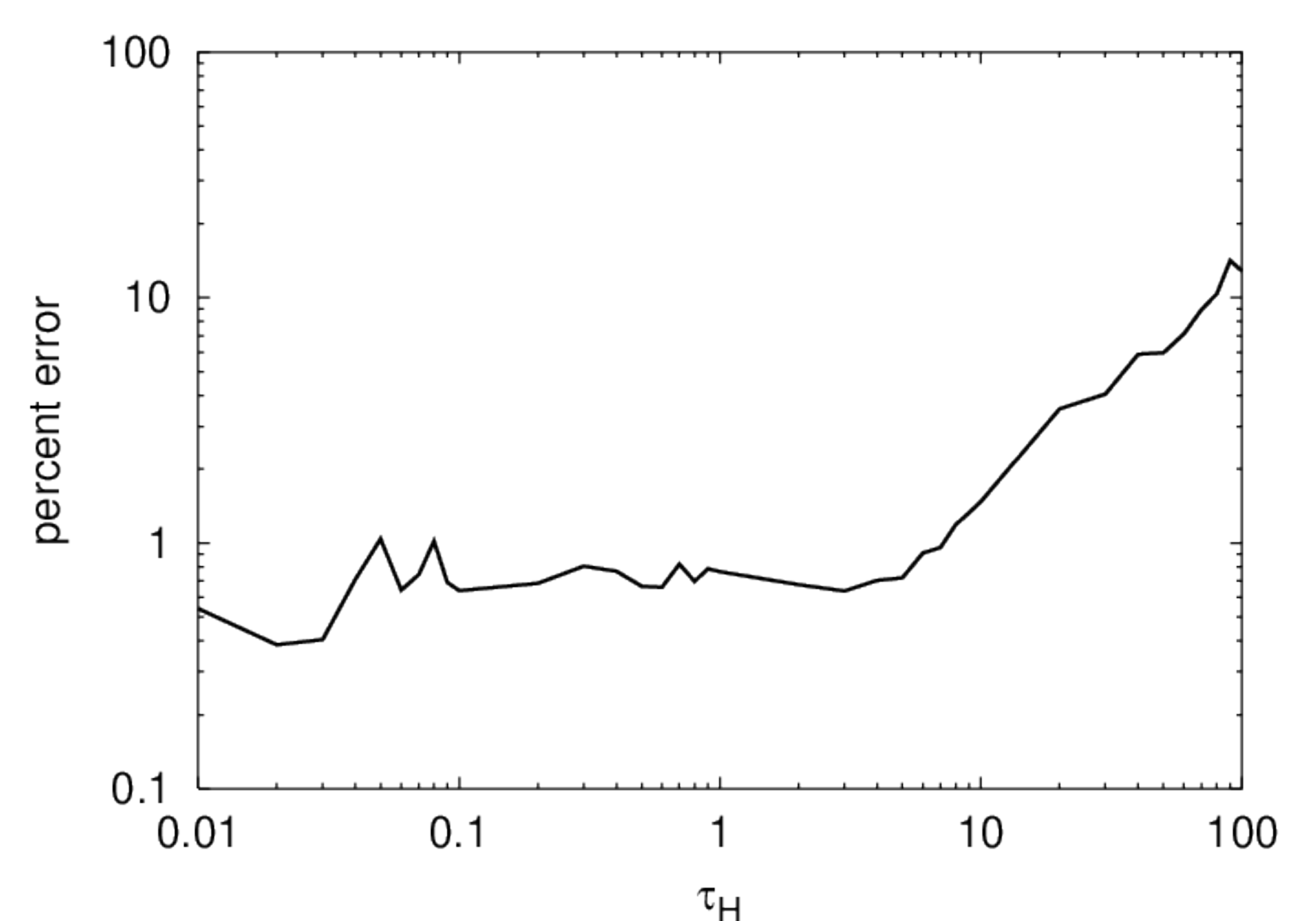,width=0.40\textwidth}}\quad
    \subfigure[Layer 10]{\epsfig{figure=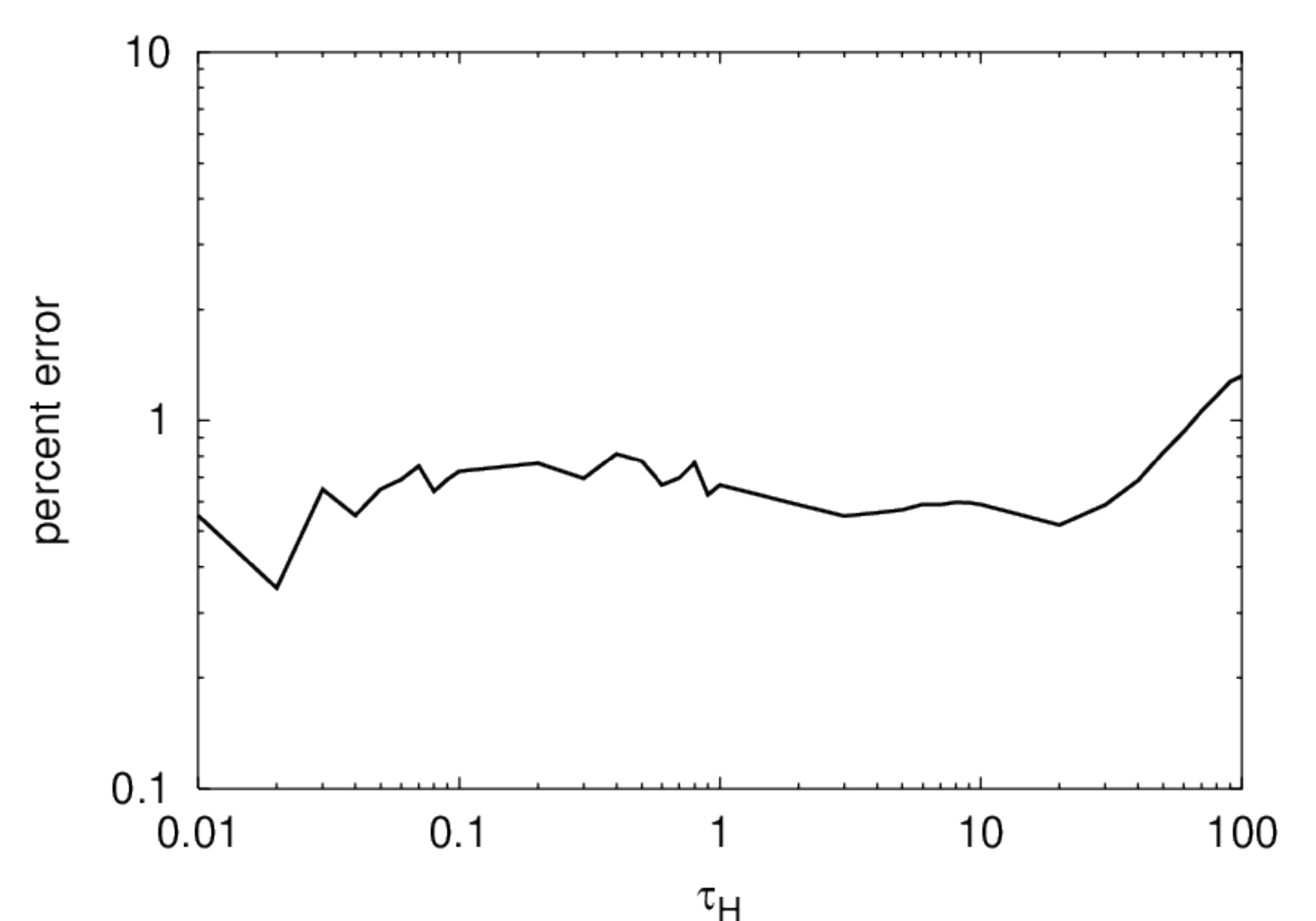,width=0.40\textwidth}}}
\caption{Same as \fig{para1} with $\omega=0.9$}
\label{fig:para3}
\end{figure}

In the preceding example a linear blackbody intensity profile was used for convergence tests concerning slab emission. This kind of blackbody intensity profile is not relevant for radiative flux divergence computations in the limit of strong optical thicknesses~: with the underlying idea of Rosseland (diffusion) approximation, the radiative budget is indeed only function of the blackbody profile second order derivative. \fig{para1}-\ref{fig:para3} therefore present convergence tests with the same slab configuration as above, but with a parabolic blackbody intensity profile($B_0$ at slab boundaries and $B_0 + \Delta B$ at slab center) : $B(z)=B_0 +\Delta B \left[1- 4\Bigl( \frac{z}{H}-\frac{1}{2}\Bigr)^{2} \right]$. Computations are performed using a slab discretization into $20$ layers of same thickness, with $N=10000$ statistical realizations per layer. Presented results are the average value of the radiative flux divergence within each layer.

\fig{para1}a-\ref{fig:para3}a display the radiative flux divergence profile for different values of the slab total optical thickness $\tau_{H}$. In these successive three figures, the single scattering albedo is respectively equal to $0.01$, $0.50$ and $0.90$. For the same values of single scattering albedo, \fig{para1}c - \ref{fig:para3}c and \fig{para1}d - \ref{fig:para3}d display radiative flux divergence averages in layers 3 and 10 respectively, as function of slab total optical thickness $\tau_{H}$. Standard deviations are presented in \fig{para1}b - \ref{fig:para3}b, \fig{para1}e - \ref{fig:para3}e and \fig{para1}f - \ref{fig:para3}f.

Results concerning layer 3 and layer 10 are presented in logarithmic scale in order to highlight the behaviors in the optically thin and optically thick limits where Monte Carlo algorithms commonly encounter convergence difficulties. In the optically thin limit, the radiative flux divergence is proportional to $k_a$, and therefore to $\tau_{H}$ (when both layer width $H$ and single scattering albedo $\omega$ are fixed). In the optically thick limit, short-distance energy redistribution processes are dominant and the radiative flux divergence follows the diffusion approximation. In the case of a parabolic blackbody intensity profile, it is constant across the slab and (for fixed values of $H$ and $\omega$) inversely proportional to $\tau_{H}$ (see \ap{A}).
Analytical results corresponding to the diffusion approximation are superimposed to the Monte Carlo results in \fig{para1}c-\ref{fig:para3}c and \fig{para1}d-\ref{fig:para3}d. Also presented are the analytical results corresponding to the pure absorption approximation (neglecting scattering)~: these analytical solutions are available, in the specific case of a parabolic blackbody intensity profile, thanks to the 4th and 5th exponential integral functions (see \ap{A}).

The results of \fig{para1} lead to the same conclusions as those of figure 7-8 in \cite{amaury02}~: for low albedoes, the convergence qualities of the present algorithm are similar to those of the previous algorithm designed for purely absorbing media \footnote{Note that a scaling error was made in \cite{amaury02}~: results of figure 8 were presented omitting to divide by a factor $25$ corresponding to the narrow band width $d \eta=25 cm^{-1}$ with which computations were held}. This is compatible with the fact that, for $\omega=0.01$, the pure absorption approximation appears as accurate for all optical thicknesses from $10^{-2}$ to $10^2$. Using 10000 statistical realizations per layer, the statistical uncertainty (more precisely the standard deviation) remains lower than a few percents for layer 10~; it reaches 10\% for layer 3 at $\tau_{H}=10$ and is independant of optical thickness above $\tau_{H}=10$. As explained in \cite{amaury02}, the fact that the uncertainty becomes independant of optical thickness at high optical thicknesses (whereas it diverges for standard Monte Carlo algorithms) comes from the fact that the boundary-based sampling of emission positions is idealy adapted to optical thickness and that the only remaining task is to perform the integration over the blackbody intensity profile, which is independant of optical thickness. The fact that higher uncertainties are observed for layer 3 than for layer 10 is due to symetry reasons~: the radiative balance of layer 10 is the sum of the net-exchanges through its bottom and top interfaces, that are of same sign, whereas the radiative balance of layer 3 is the difference between a heating and a cooling term, all net-exchanges being computed with similar uncertainties.

\fig{para2} and \fig{para3} lead to very similar observations which means that in terms of required numbers of statistical realizations, the conclusions of Sec. 3 are still valid for radiative flux divergence calculations~: no specific difficulty is encountered with the proposed algorithm up to extreme values of both absorption and scattering optical thicknesses (except for extreme cases where both optical thickness $\tau_{H}$ and single scattering albedo $\omega$ are very high, typically $\tau_{H}=100$ and $\omega=0.9999$). The average numbers of scattering events are not displayed in these figures as no additional observation can be made compared to those made in the preceding section~: it increases less rapidly with the present algorithm than with a standard Monte Carlo algorithm, which partially compensates the convergence limit at high $\tau_{H}$ and high $\omega$.

\section{Conclusion}

The above presented algorithm is an extension to scattering media of the algorithm introduced in \cite{amaury02} as a way to bypass the difficulties encountered by standard Monte Carlo algorithms in the optically thick limit. It is based on a boundary-based net-exchange formulation together with a detailed optimization of optico-geometric sampling laws. It is little sensitive to optical thickness up to both extreme values of absorption optical thickness and scattering optical thickness, two major difficulties of standard Monte Carlo algorithms. As it is based on a net-exchange formulation, it also encounters no difficulty when applied to quasi-isothermal configurations. As will be presented in a forthcoming publication, this algorithm is in particular suitable for detailed analysis of infrared radiation in the terrestrial atmosphere, in which are simultaneously encountered wide ranges of absorption optical thicknesses (because of the line spectra of atmospheric gases) and wide ranges of scattering optical thicknesses (from optically thin dust clouds to optically thick water clouds) \cite{EGS,Eurotherm}.

Structurally speaking, the proposed algorithm is very much similar to most standard Monte Carlo algorithms, except for the sampling of emission positions that is modified according to the boundary-based approach. All optimized sampling laws are also mathematically very simple and corresponding random generation procedures introduce no specific difficulty. Altogether, the proposed algorithm should therefore be easy to implement on the basis of any existing Monte Carlo code. We also hope that the presented formal derivations should allow that the reader derives its own sampling laws for best optimization in front of specific configurations.

Finally, a difficulty remains in the limit of very high scattering optical thicknesses combined with very low absorption optical thicknesses. We believe that this difficulty (that was already well identified and intensively explored for nuclear shielding applications \cite{Hammersley,Berger} ) can only be faced working on the diffusive random walk itself, using formulation efforts and sampling laws adaptations. This point was not addressed in the present paper and it will undoubtedly require further detailed analysis of the statistics of multiple scattering optical paths in finite size systems.

\appendix
\section{Appendix A: radiative flux divergence expressions at the scattering optically thin and optically thick limits.}
\label{ap:A}

\subsection{Diffusion approximation in a planes parallel configuration.}

In the case of optically thick configurations, the diffusion approximation (which is equivalent to the Rosseland approximation) may be used. The radiative flux $q_{r}(z)$ can be written as~:

\begin{equation}
q_{r}(z)=-\frac{h \nu c}{k_{a}+k_{s}} D \frac{\partial G}{\partial z}
\end{equation}

with $G(z)=\frac{1}{h \nu c}\int_{4\pi}I(z,{\bf u})d\omega({\bf u})$ the local photon density, where $I(z,{\bf u})$ is the specific intensity at altitude $z$ in direction ${\bf u}$ and $D=\frac{1}{3(1-\omega g)}$. In optically thick systems, we can make the assumption that $G(z)$, the local photon density, is equal to the equilibrium intensity at the local temperature~: $G(z)=\frac{4\pi}{h \nu c} B(z)$, with $B(z)$ the local blackbody intensity. With the assumption of a parabolic blackbody intensity profile $B(z)=B_0 +\Delta B \left[1- 4\Bigl( \frac{z}{H}-\frac{1}{2}\Bigr)^{2} \right]$, the radiative flux becomes~:

\begin{equation}
q_{r}(z)=\frac{32\pi \Delta B}{(k_{a}+k_{s})H}D\Bigl(\frac{z}{H}-\frac{1}{2}\Bigr)
\end{equation}

And its divergence is~:

\begin{equation}
div(q_{r})(z)=\frac{32\pi \Delta B}{(k_{a}+k_{s}) H^{2}}D
\label{eq:p1}
\end{equation}

Finally, the average radiative flux divergence between altitudes $z_{i-1}$ and $z_{i}$ may be written as~:

\begin{equation}
<div(q_{r})>=\frac{\int_{z_{i-1}}^{z_{i}}div(q_{r}(z))}{z_{i}-z_{i-1}}=\frac{32\pi \Delta B}{(k_{a}+k_{s}) H^{2}}D=\frac{1}{\tau_{H}}\frac{32\pi \Delta B}{ H}D
\label{eq:p12}
\end{equation}

Note that even in the optically thick limit, the diffusion approximation is not valid for the computation of the average flux divergence in the bottom and top layers (layers $1$ and $20$ in the text). The diffusion approximation is only valid far from the boundaries.

\subsection{Absorption approximation in a plane parallel configuration with black boundaries and a parabolic black intensity profile.}
\begin{figure}[h!t]
  \begin{center}
    \epsfig{figure=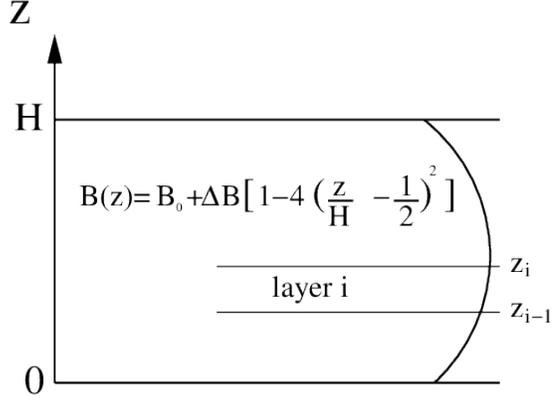 ,width=0.85\textwidth}
    \caption{Plane-parallel slab with $n$ homogeneous layers and parabolic black intensity profile.}
    \label{fig:parabol}
  \end{center}
\end{figure}

The average radiative flux divergence in layer $i$ (between altitudes $z_{i-1}$ and $z_{i}$) may be expressed as~:

\begin{equation}
<div(q_{r})>=2\pi\int_{0}^{1}\mu\Bigl(\frac{\partial I^{+}(z,\mu)}{\partial z}+\frac{\partial I^{-}(z,-\mu)}{\partial z}\Bigr)d\mu
\end{equation}
with $I^{+}(z,\mu)$ and $I^{-}(z,-\mu)$ respectively the upward and downward specific intensities at altitude $z$, in the zenithal direction $\theta$ with $\mu=cos(\theta)$. Under the pure absorption approximation, these intensities may be written as~:

\begin{equation}
I^{+}(z,\mu)=B(0)exp\bigl(-\int_{0}^{z}\frac{k_{a}(z^{\prime})}{\mu}dz^{\prime}\bigr)+\int_{0}^{z} k_{a}(z^{\prime}) B(z^{\prime}) exp\Bigl(-\int_{z^{\prime}}^{z}\frac{k_{a}(z^{\prime})}{\mu}dz^{\prime}\Bigr)\frac{dz^{\prime}}{\mu}
\end{equation}
\begin{equation}
I^{-}(z,-\mu)=B(H)exp\bigl(-\int_{H}^{z}\frac{k_{a}(z^{\prime})}{\mu}dz^{\prime}\bigr)+\int_{H}^{z} k_{a}(z^{\prime}) B(z^{\prime}) exp\Bigl(-\int_{z^{\prime}}^{z}\frac{k_{a}(z^{\prime})}{\mu}dz^{\prime}\Bigr)\frac{dz^{\prime}}{\mu}
\end{equation}

Introducing the parabolic Planck profile $B(z)=B_0 +\Delta B \left[1- 4\Bigl( \frac{z}{H}-\frac{1}{2}\Bigr)^{2} \right]$ into the above expressions leads to~:

\begin{equation}
\begin{split}
<div(q_{r})>= & \frac{2\pi}{z_{i}-z_{i-1}}\Biggl[\frac{4\Delta B}{Hk_{a}}\Biggl(E_{4}\Bigl(k_{a}z_{i}\Bigr)-E_{4}\Bigl(k_{a}z_{i-1}\Bigr)-E_{4}\Bigl(k_{a}(H-z_{i})\Bigr) \\
& +E_{4}\Bigl(k_{a}(H-z_{i-1})\Bigr)\Biggr)+\frac{8\Delta B}{(Hk_{a})^{2}}\Biggl(E_{5}\Bigl(k_{a}z_{i}\Bigr)-E_{5}\Bigl(k_{a}z_{i-1}\Bigr) \\
& -E_{5}\Bigl(k_{a}(H-z_{i})\Bigr)+E_{5}\Bigl(k_{a}(H-z_{i-1})\Bigr)\Biggr)+\frac{16\Delta B (z_{i}-z_{i-1})}{3k_{a}H^{2}}\Biggr]
\end{split}
\end{equation}

with $E_{n}$ the $n^{th}$ exponential integral~:
\begin{equation}
E_{n}(x)=\int_{0}^{1} \mu^{n-2} exp\Bigl(-\frac{x}{\mu}\Bigr)d\mu
\label{eq:En}
\end{equation}

\bibliographystyle{unsrt}

\bibliography{biblio}

\begin{thebibliography}{10}

\bibitem{amaury02}
A~De~Lataillade, J.~L. Dufresne, M.~El~Hafi, V.~Eymet, and R.~Fournier.
\newblock A net exchange monte carlo approach to radiation in optically thick
  systems.
\newblock {\em Journal of Quantitative Spectroscopy and Radiative Transfer},
  74:563--584, 2002.

\bibitem{Hammersley}
J.M. Hammersley and D.C. Handscomb.
\newblock {\em Monte-Carlo methods}.
\newblock John Wiley, New York, 1964.

\bibitem{Howell02}
J.R. Howell.
\newblock Application of monte carlo to heat transfer problems.
\newblock {\em Advances in Heat Transfer}, 5:1--54, 1969.

\bibitem{Howell01}
J.R. Howell.
\newblock The monte-carlo method in radiative heat transfer.
\newblock {\em Journal of Heat Transfer}, 120:547--560, 1998.

\bibitem{amaury01}
A.~De~Lataillade, S.~Blanco, Y.~Clergent, J.~L. Dufresne, M.~El~Hafi, and
  R.~Fournier.
\newblock Monte-carlo method and sensitivity estimations.
\newblock {\em Journal of Quantitative Spectroscopy and Radiative Transfer},
  75:529--538, 2002.

\bibitem{Walters}
D.V. Walters and R.O. Buckius.
\newblock Rigorous development for radiation.
\newblock {\em International Journal of Heat and Mass Transfer},
  35-12:3323--3333, 1992.

\bibitem{Fournier04}
M.~Cherkaoui, J.~L. Dufresne, R.~Fournier, J.~Y. Grandpeix, and A.~Lahellec.
\newblock Monte-carlo simulation of radiation in gases with a narrow-band model
  and a net-exchange formulation.
\newblock {\em ASME Journal of Heat Transfer}, 118:401--407, 1996.

\bibitem{Fournier03}
M.~Cherkaoui, J.~L. Dufresne, R.~Fournier, and J.~Y. Grandpeix.
\newblock Radiative net exchange formulation within 1d gaz enclosures with
  reflective surfaces.
\newblock {\em ASME Journal of Heat Transfer}, 120:275--278, 1998.

\bibitem{Dufresne00}
J.~L. Dufresne, R.~Fournier, and J.~Y. Grandpeix.
\newblock Méthode de monte-carlo par échanges pour le calcul des bilans
  radiatifs au sein d'une cavité 2d remplie de gaz.
\newblock {\em Compte-rendu de l'Académie des Sciences, Paris}, 326 Série II
  b:33--38, 1998.

\bibitem{Tesse01}
L.~Tessé, F.~Dupoirieux, B.~Zamuner, and J.~Taine.
\newblock Radiative transfer in real gases using reciprocal and forward monte
  carlo methods and a correlated-k approach.
\newblock {\em International Journal of Heat and Mass Transfer}, 3, issue
  13:2797--2814, 2002.

\bibitem{Martin}
W.R. Martin and G.C. Pomraning.
\newblock Monte carlo analysis of the backscattering of radiation from a sphere
  to a plane.
\newblock {\em Journal of Quantitative Spectroscopy and Radiative Transfer}, 43
  - 2:115--126, 1990.

\bibitem{Dufresne01}
J.-L. Dufresne, R.~Fournier, and J.-Y. Grandpeix.
\newblock Inverse gaussian k-distributions.
\newblock {\em Journal of Quantitative Spectroscopy and Radiative Transfer}, 61
  n4:433--441, 1999.

\bibitem{Green}
J.~S.~A. Green.
\newblock Division of radiative streams into internal transfer and cooling to
  space.
\newblock {\em Quarterly Journal of the Royal Meteorological Society},
  93:371--372, 1967.

\bibitem{case}
K.M. Case and P.F. Zweifel.
\newblock {\em Linear Transport Theory}.
\newblock Addison-Wesley Publishing Company, 1967.

\bibitem{Blanco01}
S.~Blanco and R.~Fournier.
\newblock An invariance property of diffusive random walks.
\newblock {\em Europhysics Letter}, 61 (2):168--173, 2003.

\bibitem{Feller}
W.~Feller.
\newblock {\em An introduction to probability theory and its applications, 2nd
  edition}, volume~1.
\newblock John Willey and Sons, New York, 1966.

\bibitem{EGS}
V.~Eymet, S.~Blanco, R.~Fournier, and J.L. Dufresne.
\newblock Longwave radiative exchange analysis of cloudy atmospheres with a net
  exchange formulation .
\newblock In {\em EGS-AGU-EUG, joint Assembly}, Nice, France, 06-11 April 2003.

\bibitem{Eurotherm}
V.~Eymet, S.~Blanco, R.~Fournier, and J.L. Dufresne.
\newblock A monte carlo method to develop radiative transfer parametrizations
  for terrestrial gcm.
\newblock In {\em Proceedings of Eurotherm 73 on Computational Thermal
  Radiation in Participating Media}, pages 139--148, Mons, Belgium, 15-17 April
  2003.

\bibitem{Berger}
M.J. Berger and J.~Dogget.
\newblock Reflection and transmission of gamma radiation by barriers :
  semianalytic monte carlo calculation.
\newblock {\em J. Res. Net. Bur. Stand.}, 56:89--98, 1956.

\end{thebibliography}
\end{document}